\documentclass[twocolumn,showpacs,preprintnumbers,amsmath,amssymb]{revtex4}

\usepackage{graphicx}
\usepackage{dcolumn}
\usepackage{bm}

\begin{document}

\input epsf.sty



\title{Typical state of an isolated quantum system with fixed energy and unrestricted participation of eigenstates}

\author{Boris V. Fine}

\affiliation{ Institute for Theoretical Physics, University of Heidelberg, Philosophenweg 19, 69120 Heidelberg, Germany }

\date{March 2, 2009}

\begin{abstract}
This work describes the statistics  for the occupation numbers of quantum levels in a large isolated quantum system, where all possible superpositions of eigenstates are allowed, provided all these superpositions have the same fixed energy. Such a condition is not equivalent to the conventional microcanonical condition, because the latter limits the participating eigenstates to a very narrow energy window. The statistics is obtained analytically for both the entire system and its small subsystem. In a significant departure from the Boltzmann-Gibbs statistics, the average occupation numbers of quantum states exhibit in the present case weak algebraic dependence on energy. In the macroscopic limit, this dependence is routinely accompanied by the condensation into the lowest energy quantum state. 
This work contains initial numerical tests of the above statistics for finite systems, and also reports the following numerical finding: When the basis states of large but finite random matrix Hamiltonians are expanded in terms of eigenstates, the participation of eigenstates in such an expansion obeys the newly obtained statistics.
The above statistics might be observable in small quantum systems, but for the macroscopic systems, it rather reenforces doubts about self-sufficiency of non-relativistic quantum mechanics for justifying the Boltzmann-Gibbs equilibrium.
\end{abstract}
\pacs{}


\maketitle

\narrowtext
\pagebreak

\section{Introduction}

It is known empirically since the introduction of Quantum Hypothesis by Planck that thermal Boltzmann-Gibbs distribution works extremely well for quantum systems. However, purely quantum derivation of this distribution is still not on a satisfactory ground. In a self-contained derivation one should be able  to start from a large isolated system and then obtain the statistical distribution for a small subsystem. The conventional derivation of the Boltzmann-Gibbs distribution proceeds by postulating the micro-canonical condition. This condition has different status in classical and quantum mechanics. In classical mechanics the micro-canonical condition rests on the further assumption of equipartition on the constant energy shell in the phase space, which in turn can be justified by the dynamical chaos caused by the non-linear interactions between particles. In this respect, the classical derivation is in better shape. In contrast, the quantum systems have no phase space, and also they are fundamentally linear.  A typical state of an isolated quantum system is not an eigenstate but a superposition of eigenstates:
\begin{equation}
\Psi = \sum_{i=1}^{N} C_i \phi_i,
\label{Psi}
\end{equation}
where $\Psi$ is the wave function of the superposition, $\phi_i$ is the wave function of the $i$-th eigenstate, $C_i$ the corresponding complex amplitude, and $N$ the total number of eigenstates.
Therefore, the straightforward  counterpart of the classical microcanonical condition would be to constrain the possible choices of $\Psi$ to the ``energy shell'' in the Hilbert space: 
\begin{equation}
\sum_{i=1}^{N}  E_i p_i  = E_{\hbox{\scriptsize av}},
\label{epsav}
\end{equation}
where $p_i = |C_i|^2$ are the occupation numbers of quantum states and $E_{\hbox{\scriptsize av}}$ is the energy of the quantum superposition set externally and referred to below as ``average energy.'' 

Condition (\ref{epsav}) is, however, different from the conventional microcanonical condition, because  the latter involves the summation only over the eigenstates inside a very small energy window  $E_{\hbox{\scriptsize av}} \pm \delta E$. 

Why the system should limit itself to a small energy window is difficult to justify  unless, for example, one assumes that the quantum system is subjected to an external source of decoherence with the subsequent collapse of the density matrix. However, the introduction of collapse would imply that non-relativistic quantum mechanics is not self-contained, when it comes to justifying the Boltzmann-Gibbs equilibrium. 

Besides the conceptual issues, there are also practical ones. When a well isolated quantum system having not too many particles but many quantum levels is shaken in an experiment and then left to itself, the energy window of the participating eigenstates can easily become larger than the temperature. Would such a system end up exhibiting Boltzmann-Gibbs statistics?

It is clear {\it a priori}, that a significant departure from the narrow-energy-window constraint can easily lead to deviations from the Boltzmann-Gibbs statistics\cite{Aarts-00}. One can, consider, for example the case of two narrow energy windows. Still one can hope that somehow the ``most probable departure'' from the narrow energy window condition would still support the Boltzmann-Gibbs equilibrium.

If one is to begin addressing the above issues, the unavoidable limit to consider is the system of $N >> 1$ quantum levels with constraint (\ref{epsav}) and no limit on the energy window, i.e. all quantum superpositions of form (\ref{Psi}) satisfying condition (\ref{epsav}) are equally probable and, therefore, the probability density as a function of the complex amplitudes $C_i$ is proportional to the volume element on a manifold in the Hilbert space constrained by Eq.(\ref{epsav}). Following Ref.\cite{Bender-05}, I call  this condition ``quantum micro-canonical'' (QMC) to contrast it with the conventional micro-canonical condition. 

The general approach of assigning the probability on the basis of volume in Hilbert space has received a good degree of attention in recent years. Some of the relevant works\cite{Gemmer-04,Popescu-06,Goldstein-06,Reimann-08} applied this approach to the conventional micro-canonical case with small energy window for the participating eigenstates. Other works\cite{Brody-05,Bender-05,Naudts-06,Brody-07}, however, have looked precisely at the QMC alternative. 

In particular, it was found in Ref.\cite{Bender-05} for the case of equally spaced levels (and confirmed in the present work for the general case) that, as $N \rightarrow \infty$, the volume of the Hilbert space as function of $E_{\hbox{\scriptsize av}}$ acquires the character of a $\delta$-function with peak located at $E_{\hbox{\scriptsize av}} = {1 \over N} \sum_i E_i$. This result, however, does not imply that it is not important to consider the case of $E_{\hbox{\scriptsize av}}$ different from the above value. The situation here is analogous to the conventional micro-canonical description, when the most probable position of the narrow energy window would correspond to the infinite temperature, but one would still like to know the result for a finite temperature.

The goal of the present work is to obtain from the QMC condition the statistics for the occupation numbers of individual quantum states both for the entire isolated quantum system (Section~\ref{whole}), and for the density matrix of a small part of it (Section~\ref{subsystem}). It is to be shown analytically that this statistics is dramatically different from the Boltzmann-Gibbs statistics: the occupation numbers of quantum states decay with energy algebraically rather than exponentially, and, in addition, the macroscopic limit is routinely accompanied by condensation into the lowest energy state. 

Section~\ref{numerical} contains preliminary numerical tests of some of the above analytical results for finite systems. This section also reports a numerical finding that the expansion of the basis states of large but finite random matrix Hamiltonians in terms of the eigenstates of these Hamiltonians obeys the statistics found in this work.

The implications of the above results are to be discussed in the concluding remarks(Section~\ref{concluding}). 

The notion of chaos does not play any role in the forthcoming derivation, but it will also be touched briefly in the concluding remarks.

Even though the formal treatment below is based essentially on the geometrical analysis of many-dimensional manifolds, it should not escape the readers that the end result is similar to the Boltzmann-Gibbs distribution for grand canonical ensemble --- though  not for the average values of the occupation numbers $\langle p_i \rangle$, which the Boltzmann-Gibbs distribution aims at predicting, but rather for the probabilities of variable $p_i$ to admit different values.  In other words, it is an example of superstatistics\cite{Beck-03} --- consequence of the fact that the occupation numbers $p_i$, which are interpreted as quantum probabilities, are themselves subject to the probability distribution. 

\section{Statistics for the eigenstates of an isolated quantum system}
\label{whole}

\subsection{Formulation of the problem}
\label{form1}

The Hilbert space is parameterizable by the absolute values $|C_i|$ and the phases $\varphi_i$ of the complex amplitudes $C_i = |C_i| e^{i \varphi_i}$. Without the energy and the normalization constraints, the volume element in this space is given by $dV = \prod_i^N |C_i| \  d \varphi_i \  d |C_i| = 
{1 \over 2} \prod_i^N  d \varphi_i \  d (|C_i|^2)$. Both energy and the normalization constraints involve only $|C_i|^2$, and, therefore, when assigning the probabilities in the Hilbert space, one can integrate over phases $\varphi_i$ and then deal only with the subspace of occupation numbers 
\begin{equation}
 p_i = |C_i|^2.
\label{pi}
\end{equation}
The Hilbert space volume element is then given by 
\begin{equation}
 dV = \prod_i^N dp_i
\label{dV}
\end{equation}
with an unimportant prefactor.
It is always to be assumed that $N>>1$. 

In the rest of this paper, I will be calculating the volumes of manifolds in the Euclidean space of variables $p_i$ constrained by

(i) energy condition (\ref{epsav});

(ii) normalization condition
\begin{equation}
\sum_i^N p_i = 1;
\label{norm}
\end{equation}
and 
(iii) positivity condition
\begin{equation}
p_i \geq 0,\ \ \ \ \  \forall i.
\label{positiv}
\end{equation}

Energies $E_i$ are ordered by their values with the minimum one being $E_{\hbox{\scriptsize min}} \equiv E_1$ and the maximum one $E_{\hbox{\scriptsize max}} \equiv E_N$. The zero reference point for the energies is chosen such that
\begin{equation}
\sum_i^N E_i = 0.
\label{epsref}
\end{equation}
When not stated otherwise, it will be assumed below that
\begin{equation}
E_{\hbox{\scriptsize av}} < 0.
\label{eavless}
\end{equation}

For a given value of index $k$, I denote the $(N-1)$-dimensional Euclidean space of all variables $\{p_i\}$ with $i \neq k$ as  $\{p_i\}_k$.

The probability of the $k$-th occupation number to have certain value $p_k$ is proportional to the volume $V_k(p_k)$ of the (N-3)-dimensional manifold in the $\{p_i\}_k$-space --- to be denoted as $M_k$ --- constrained by conditions (\ref{positiv})  in combination with
\begin{equation}
\sum_{i, i\neq k}^N p_i = 1 - p_k,
\label{normK}
\end{equation}
---consequence of (\ref{norm}), and
\begin{equation}
\sum_{i, i\neq k}^{N}  (E_i - E_{\hbox{\scriptsize av}})  p_i   =  - (E_k - E_{\hbox{\scriptsize av}}) p_k.
\label{epsavK0}
\end{equation}
Condition (\ref{epsavK0}), while obviously originating from (\ref{epsav}), requires a preliminary manipulation equivalent to shifting the origin of the energy axis to $E_{\hbox{\scriptsize av}}$. Namely, $E_{\hbox{\scriptsize av}}$ in the right-hand side of Eq.(\ref{epsav}) has to be multiplied by $\sum_{i}^{N} p_i$ [equal to 1 according to (\ref{norm})] and then the result transformed to Eq.(\ref{epsavK0}). Important for the subsequent derivation is the fact that energy hyperplane represented by Eq.(\ref{epsavK0}) crosses the origin of the $\{p_i\}_k$-space, when $p_k = 0$.

The manifold $M_k$ has a character of $(N-3)$-dimensional polygon with flat faces, edges, etc., because all conditions constraining it represent hyperplanes in the $\{p_i\}_k$-space.

The probability distribution of $p_k$ is then
\mbox{$P(p_k) = V_k(p_k)/\left(\int_0^1 V_k(p_k^{\prime})dp_k^{\prime}\right)$}, and the average value of $p_k$ is 
\begin{equation}
\langle p_k \rangle = {
\int_0^1 p_k^{\prime} V_k(p_k^{\prime})dp_k^{\prime} 
\over 
\int_0^1 V_k(p_k^{\prime})dp_k^{\prime} }.
\label{pkav}
\end{equation}

\subsection{ The case of  $p_k \ll 1$}
\label{pk-less-1}

I first consider the case $p_k \ll 1$.

The manifold constrained by conditions (\ref{positiv}, \ref{normK}, \ref{epsavK0}) can now be described as follows:

The intersection of (\ref{positiv}) and (\ref{normK}) is a many-dimensional analog of a tetrahedron. It has dimension $(N-2)$ --- equal to that of the normalization hyperplane (\ref{normK}) with  $(N-1)$ vertices located at the intersections of the $(N-1)$ axes of the $\{p_i\}_k$ space with the hyperplane (\ref{normK}),
i.e. in the $\{p_i\}_k$-space, each of the vertices has coordinates of type $(0, 0, ..., 1-p_k, ...,0)$ --- all projections are zero, except for one, which is equal to $1-p_k$. I call the resulting object ``Hypertetrahedron''. This Hypertetrahedron is then cross-sected by the energy hyperplane (\ref{epsavK0}).

When $p_k$ is small,  the renormalization of the volume $V_k(p_k)$ with respect to $V_k(0)$ can be decomposed into the ``normalization factor'' $F_N$ due to the non-zero value of $p_k$ in Eq.(\ref{normK}) and the ``energy factor'' $F_E$ due to the non-zero value of $p_k$ in Eq.(\ref{epsavK0}):
\begin{equation}
V_k(p_k) = V_k(0) F_N F_E .
\label{renorm}
\end{equation}

The normalization factor is given exactly by 
\begin{equation}
F_N = (1-p_k)^{N-3}
\label{normfactor}
\end{equation}
for large or small $p_k$. It is the consequence of the fact that the change of  $(1-p_k)$ in the right-hand side of Eq.(\ref{normK}) rescales the distance between any point of the Hypertetrahedron and the origin of 
the $\{p_i\}_k$ space by factor $(1-p_k)$. Since the energy hyperplane (\ref{epsavK0}) passes through the origin (at $p_k = 0$), each of the $(N-3)$ dimensions of the intersection manifold simply undergoes rescaling by factor $(1-p_k)$ thus leading to factor (\ref{normfactor}).

The calculation of the energy factor $F_E$ requires more effort. The change of $ - (E_k - E_{\hbox{\scriptsize av}}) p_k$ in the right-hand side of Eq.(\ref{epsavK0}) shifts the energy hyperplane in the transverse direction, but the resulting change of manifold $M_k$ does not any longer amount to a self-similar rescaling.

The volume $V_k$ of the $(N-3)$-dimensional manifold $M_k$ can in general be presented as a product of $(N-3)$ characteristic linear parameters $\eta_{k \alpha}$:
\begin{equation}
V_k = \prod_{\alpha=1}^{N-3} \eta_{k \alpha}
\label{etas}
\end{equation}
These parameters can be defined iteratively in the following way: 
$\eta_{k 1} = V_k/V_{k, N-4}$, where $V_{k, N-4}$ is the volume of one of the
$(N-4)$-dimensional faces of $M_k$; $\eta_{k 2} = V_{k, N-4}/V_{k, N-5}$, where
where $V_{k, N-5}$ is the volume of one of the $(N-5)$-dimensional faces of the $(N-4)$-dimensional face selected in the previous step; etc.

After the small shift of the energy hyperplane (\ref{epsavK0}) by 
$-(E_k - E_{\hbox{\scriptsize av}}) p_k$, each linear parameter $\eta_{k \alpha}$ changes slightly to 
\begin{equation}
\eta_{k \alpha}(p_k) = \eta_{k \alpha}(0) [1- \lambda_{k \alpha} (E_k - E_{\hbox{\scriptsize av}}) p_k], 
\label{etapk}
\end{equation}
where $\lambda_{k \alpha}$ are unknown rescaling coefficients. 
As a result,
\begin{equation}
F_E = \prod_{\alpha=1}^{N-3} [1- \lambda_{k \alpha} (E_k - E_{\hbox{\scriptsize av}}) p_k]
\approx e^{- (N-3) \lambda_k (E_k - E_{\hbox{\scriptsize av}}) p_k },
\label{FE}
\end{equation}
where 
\begin{equation}
\lambda_k = {1 \over N-3} \sum_{\alpha}^{N-3} \lambda_{k \alpha}.
\label{lambdak}
\end{equation}

Coefficients $\lambda_{k \alpha}$ are not well differentiable with respect to $p_k$ and $E_{\hbox{\scriptsize av}}$,
because the change of $p_k$ and $E_{\hbox{\scriptsize av}}$ is accompanied by the change in the number of vertices of manifold $M_k$. However, the internal self-consistency of the present treatment indicates, that the overall renormalization factor $F_E$ depends on $p_k$ and $E_{\hbox{\scriptsize av}}$ sufficiently weakly and can be efficiently approximated.

Even though each renormalization factor $[1- \lambda_{k \alpha} (E_k -E_{\hbox{\scriptsize av}}) p_k]$ in Eq.(\ref{FE} is very close to 1, the product of the $(N-3)$ of these factors may be significantly smaller than 1 without compromising the validity of small-$p_k$ approximation (\ref{FE}) for $F_E$ alone.  Yet, when $p_k \ll 1$, but both $F_N$ and $F_E$ depart significantly from one,
one can worry that the effects of shifting the normalization  and the energy hyperplanes [(\ref{normK}) and (\ref{epsavK0})] do not commute with each other, and therefore, the resulting renormalization is not equal to the product of $F_N$ and $F_E$. This is, however, not the case, because the shift of the normalization hyperplane amounts to a simple rescaling, and after that, the shift of the energy hyperplane always begins from the manifold of the same geometry.

Central to the present work is the result that in the leading order in $1/N$, $\lambda_k$ is simply independent of $k$. I denote this independent value as $\lambda$ without a subscript. It is shown in Appendix~\ref{central} that the linear parameters in (\ref{etas}) can always be chosen such that all but one summands are equal to each other in the expressions for two different renormalization coefficients $\lambda_k = {1 \over N-3} \sum_{\alpha}^{N-3} \lambda_{k \alpha}$
and $\lambda_l = {1 \over N-3} \sum_{\alpha}^{N-3} \lambda_{l \alpha}$, i.e.
$\lambda_{k \alpha} = \lambda_{l \alpha}$ for all $\alpha$ except for one value $\alpha_0$. In a typical case, however, $\lambda_{k \alpha_0}$ and $\lambda_{l \alpha_0}$ are much smaller than the rest of their respective sums.

Substituting $\lambda$ instead of $\lambda_k$ in  (\ref{FE}) and then combining in (\ref{renorm}) the resulting expression for $F_E$ with $F_N$ from (\ref{normfactor}) while keeping only the leading order in $N$, I obtain
\begin{equation}
V_k(p_k) = V_k(0) e^{-N p_k[1 + \lambda(E_k - E_{\hbox{\scriptsize av}})]},
\label{Vkpk1}
\end{equation}
As long as 
\begin{equation}
1 + \lambda(E_k - E_{\hbox{\scriptsize av}}) \gg {1 \over N},
\label{small-pk}
\end{equation}
$V_k(p_k)$ decays almost completely, when $p_k \ll 1$, and, therefore,
expression (\ref{Vkpk1}) is sufficient to calculate 
$\langle p_k \rangle$ from Eq.(\ref{pkav}), which gives
\begin{equation}
\langle p_k \rangle = {1 \over N [1 + \lambda (E_k - E_{\hbox{\scriptsize av}})]}.
\label{pav2}
\end{equation}
The value of $\lambda$ can now be found numerically from either of the following two conditions  originating, respectively, from Eqs.(\ref{norm}) and (\ref{epsav}):
\begin{equation}
\sum_{k=1}^N \langle p_k \rangle = 1,
\label{normav}
\end{equation}
or
\begin{equation}
\sum_{k=1}^N  (E_k - E_{\hbox{\scriptsize av}}) \langle p_k \rangle   =  0.
\label{epsavav}
\end{equation}
Expression (\ref{pav2}) for $\langle p_k \rangle$ has the property that, if the value of $\lambda$ is found from one of the two conditions --- (\ref{normav}) or (\ref{epsavav}), the other one is fulfilled automatically.

The value of $\lambda$ thus obtained becomes a function of average energy
$\lambda[E_{\hbox{\scriptsize av}}]$. Below I use $\lambda$ both with and without its argument. In order to distinguish the argument of function $\lambda[E_{\hbox{\scriptsize av}}]$ from the multiplication of $\lambda$ by an expression in parentheses, the argument of $\lambda[E_{\hbox{\scriptsize av}}]$, if present, will always follow $\lambda$ in square brackets. 

It is useful to present the conditions (\ref{normav}, \ref{epsavav}) also in the integral form with the values of $\langle p_k \rangle$ substituted from (\ref{pav2}):
\begin{equation}
{1 \over N} \int_{- \infty}^{+ \infty} 
{
\nu(E) d E
\over
1 + \lambda (E - E_{\hbox{\scriptsize av}})
}
= 1;
\label{nunorm}
\end{equation}
\begin{equation}
\int_{- \infty}^{+ \infty} 
{
(E - E_{\hbox{\scriptsize av}}) \nu(E) d E
\over
1 + \lambda (E - E_{\hbox{\scriptsize av}})
}
= 0.
\label{nueav}
\end{equation}
where ($\nu(E)$ is the density of states corresponding to the energy spectrum $\{ E_k \}$ and satisfying the condition $\int_{- \infty}^{+ \infty} \nu(E) = N$. 

\subsection{Meaning of $\lambda$}
\label{sub-l}

The parameter $\lambda$ or, more precisely, $N \lambda$ has the meaning of inverse Hilbert space temperature. It was introduced to describe the volume change of manifold $M_k$ in the $(N-1)$-dimensional $\{p_i\}_k$-space in response to the change in the right-hand side of the energy constraint, but, in the leading order in $1/N$,  it also describes the change of volume $V_{\hbox{\scriptsize tot}}$ of the entire energy manifold constrained by conditions (\ref{epsav},\ref{norm},\ref{positiv}) in the full $N$-dimensional Hilbert space of the problem as a function of $E_{\hbox{\scriptsize av}}$:
\begin{equation}
\lambda[E_{\hbox{\scriptsize av}}] = {1 \over N} {\partial  \over \partial E_{\hbox{\scriptsize av}} } \hbox{log} V_{\hbox{\scriptsize tot}}(E_{\hbox{\scriptsize av}}) + O(1/N)
\label{invtemp}
\end{equation}
[The extra dimension  would introduce only one extra linear parameter $\eta_0$ and one more coefficient $\lambda_{0, \alpha}$ in the sum of $(N-1)$ other comparable coefficients in the early proof that justified the single value of 
$\lambda $ for all $\{p_i\}_k$-spaces---see Section~\ref{pk-less-1} and Appendix~\ref{central}.] As a consequence,
\begin{equation}
V_{\hbox{\scriptsize tot}}(E_{\hbox{\scriptsize av}}) = V_{\hbox{\scriptsize max}} \hbox{exp} \left[ 
N \int_0^{E_{\hbox{\scriptsize av}}} \lambda[E] d E
\right]. 
\label{Vtot}
\end{equation}
where $V_{\hbox{\scriptsize max}}$ is the maximum value of $V_{\hbox{\scriptsize tot}}$ corresponding to $E_{\hbox{\scriptsize av}}=0$, which, in turn, is the average value of all energies in the spectrum as defined by Eq.(\ref{epsref}). 

In order to prove that the maximum of $V_{\hbox{\scriptsize tot}}$ is indeed located at $E_{\hbox{\scriptsize av}}=0$, one should note that, according to Eq.(\ref{invtemp}), this maximum implies $\lambda = 0$. Equation (\ref{pav2}) then gives $\langle p_k \rangle = 1/N$, which, according to Eqs.(\ref{epsref},\ref{epsavav}), can only be the case, when $E_{\hbox{\scriptsize av}}=0$. This general result is in agreement with the analysis of Refs.\cite{Bender-05}.

In principle, the point $E_{\hbox{\scriptsize av}}=0$ may or may not coincide with the maximum of $\nu(E)$, which is already a significant departure from the conventional statistics predicting the most probable state of the system (zero inverse temperature) always at the maximum of $\nu(E)$.  

Another important difference is that even when the maxima of $V_{\hbox{\scriptsize tot}}(E)$  and $\nu(E)$ coincide, $V_{\hbox{\scriptsize tot}}(E)$ decays exponentially faster than $\nu(E)$, which has the consequence that the small-$p_k$ condition (\ref{small-pk}) can be easily violated for the low-lying levels leading to a sort of condensation. 

\begin{figure} \setlength{\unitlength}{0.1cm}

\begin{picture}(100, 55)
{ 
\put(0, 0){ \epsfxsize= 3.2in \epsfbox{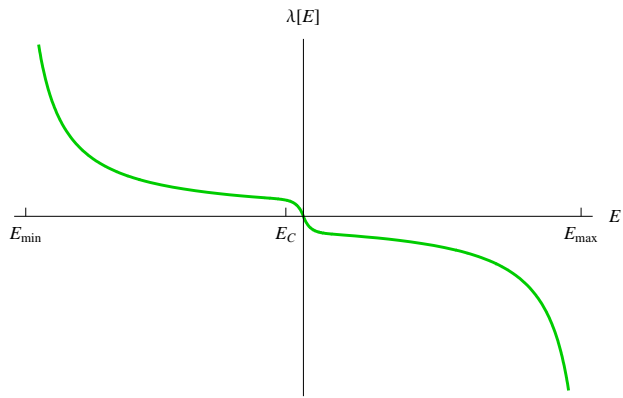} }
}
\end{picture} 
\caption{(Color online) Sketch of $\lambda[E]$. Here $E$ is the average energy identical with $E_{\hbox{\scriptsize av}}$ in the right-hand-side of Eq.(\ref{epsav}). Critical average energy $E_C$ is defined by Eq.(\ref{ElEC}).
} 
\label{fig-lambda} 
\end{figure}

A typical dependence of parameter $\lambda$ on the value of the average energy is sketched in Fig.~\ref{fig-lambda}. One should, in particular, remember that $E_{\hbox{\scriptsize av}} < 0$ (the default assumption for most of this paper) corresponds to $\lambda > 0$ and {\it vice versa}.


\subsection{Beyond small $p_k$}
\label{beyond}

In general, whether or not the preceeding description is sufficient for calculating the occupation of all quantum levels depends on the spectrum of the problem and on the value of $E_{\hbox{\scriptsize av}}$. One should attempt to solve the above system of equations and see whether $\langle p_k \rangle \ll 1$ for all $k$. If not, then one should use the following results extended to the case 
of $\langle p_k \rangle \sim 1$.

Expression (\ref{pav2}) has a pole as a function of $E_k$, which I denote as $E_{\lambda}$:
\begin{equation}
 E_{\lambda} = E_{\hbox{\scriptsize av}} - {1 \over \lambda}.
\label{elambda}
\end{equation}
Parameter $E_{\lambda}$ in the present treatment is analogous to the chemical potential in conventional thermodynamics. Like $\lambda[E_{\hbox{\scriptsize av}}]$, it is the function of the average energy, $E_{\lambda}[E_{\hbox{\scriptsize av}}]$, and, likewise, I will be using square brackets to refer to the argument of this function.

Condition (\ref{small-pk}) is satisfied for all levels, when
\begin{equation}
 E_{\hbox{\scriptsize min}} - E_{\lambda} >> { E_{\hbox{\scriptsize av}} - E_{\lambda} \over N},
\label{emin-cond}
\end{equation}
which for all practical purposes translates into
\begin{equation}
 E_{\lambda} < E_{\hbox{\scriptsize min}} - O(1/N)
\label{em-el}
\end{equation}
When $\lambda$ is positive and small, it corresponds to negative $E_{\hbox{\scriptsize av}}$ sufficiently close to zero, and therefore, the preceeding solution is valid.  
However, it always becomes violated as soon as $E_{\hbox{\scriptsize av}}$ departs significantly from zero. Below I will be using variable $E_C$ to refer to the critical value of $E_{\hbox{\scriptsize av}}$ corresponding to:
\begin{equation}
 E_{\lambda}[E_C] = E_{\hbox{\scriptsize min}}.
\label{ElEC}
\end{equation}
(see Fig.~\ref{fig-lambda})

In order to understand the regime of large $p_k$, it is necessary to appreciate, that once condition (\ref{small-pk}) or (\ref{emin-cond}) is violated,  $\lambda$ would continue to describe the response of $V_k(p_k)$ to the small shift of the energy hyperplane (\ref{epsavK0}) around $p_k = 0$. However, the whole function $V_k(p_k)$ given by Eq.(\ref{Vkpk1}) either decays too slowly or increases, and, therefore, the linear approximation for the power in (\ref{Vkpk1}) becomes insufficient and Eq.(\ref{pav2}) is not justified any longer.

It is, however, shown in Appendix~\ref{integral} that the derivation of (\ref{Vkpk1}) is amenable to the case of arbitrary $p_k$. The result is
\begin{widetext}
\begin{equation}
 V_k(p_k) = V_k(0) \hbox{exp} \left\{
(N-3)
\left[ \hbox{log}(1-p_k) + \int_{E_{\hbox{\scriptsize av}}}^{E_{\hbox{\scriptsize av}} - {(E_k -E_{\hbox{\scriptsize av}}) p_k \over 1-p_k} } \lambda[E] d E
\right]
\right\}
\label{Vkpk2}
\end{equation}
\end{widetext}
where $\lambda[E]$ is to be determined  self-consistently by solving Eq.(\ref{normav}) or (\ref{epsavav}) , though this time not just for a single value of $E_{\hbox{\scriptsize av}}$ but rather for $E_{\hbox{\scriptsize av}}$ spanning most of the  allowed interval $[E_{\hbox{\scriptsize min}}, E_{\hbox{\scriptsize max}}]$ as required by the integral in (\ref{Vkpk2}). A possible algorithm of the overall self-consistent solution in discussed in Appendix~\ref{integral}. I chose not to approximate $(N-3)$ with $N$ in Eq.(\ref{Vkpk2}), because it appears to be an important correction for finite-$N$ systems.

It is possible to anticipate the outcome of the above self-consistent solution qualitatively. Once the condition (\ref{em-el}) is violated, the average occupation number of the levels affected becomes a significant fraction of one. The overall normalization constraint (\ref{normav}) then implies that the number of these exceptional levels should be small, certainly much smaller than $N$. Therefore, as $E_{\hbox{\scriptsize av}}$ continues decreasing beyond $E_C$, $E_{\lambda}$ may increase above $E_{\hbox{\scriptsize min}}$ but will stay very close to $E_{\hbox{\scriptsize min}}$. This leads to an important approximation, namely: for $E_{\hbox{\scriptsize av}} < E_C$,
\begin{equation}
 \lambda[E_{\hbox{\scriptsize av}}] \approx {1 \over E_{\hbox{\scriptsize av}} - E_{\hbox{\scriptsize min}} },
\label{lambdaA}
\end{equation}
which one obtains by substituting $E_{\lambda} \approx E_{\hbox{\scriptsize min}}$ into Eq.(\ref{elambda}).

The value of $|E_C|$ is comparable to $|E_{\hbox{\scriptsize min}}|$, when $N$ is large, but not exponentially large, as might be the case in the numerical studies (see below) and also in nano-sized systems having not too large number of particles, but very large number of levels. In these cases, the full calculation of the resulting statistics has to be done numerically.

However, for a physical system, having macroscopic number of weakly interacting components, further progress can be made analytically.

\subsection{Macroscopic system with non-degenerate ground state}
\label{macro}

\subsubsection{Definitions and assumptions about the macroscopic system}
\label{macro-def}

As macroscopic, I understand a system consisting of macroscopic number $N_s \sim 10^{23} $ of relatively weakly interacting parts. In a gas, one molecule would constitute such one part. In condensed matter systems with finite-range interactions, a part would imply a cluster of atoms, whose volume energy is much greater than the surface energy. Each part is assumed to be characterized
by a finite Hilbert space. (The limit of large Hilbert space per constituent part is considered in the next subsection.) The total number of levels in such a macroscopic system satisfies inequality:
\begin{equation}
 N > 2^{N_s} \gg N_s.
\label{2Ns-m}
\end{equation}
Due to the large number of weakly interacting parts, the density of states $\nu(E)$ of this macroscopic system is assumed to have narrow Gaussian peak around $E=0$, which is set by Eq.(\ref{epsref}) to be equal to the average of all energies $\{E_i \}$.  The mean-squared deviation of $\nu(E)$ from the above peak position is  
\begin{equation}
 \sigma_s \equiv {1 \over N} \sum_{i=1}^N E_i^2 \approx N_s \sigma_0,
\label{sigmas-m}
\end{equation}
where $\sigma_0$ is the typical mean-squared deviation for a constituent part.

The average energy for each constituent part can also be set at zero.
With the above convention, 
\begin{equation}
 |E_{\hbox{\scriptsize min}}| \sim N_s E_{\hbox{\scriptsize min} 0} \gg \sqrt{\sigma_s}.
\label{Emin-sigmas}
\end{equation}
where, $E_{\hbox{\scriptsize min} 0}$ is the typical minimum energy for a constituent part.

In the conventional micro-canonical formulation of statistical physics, the temperature $T$ corresponding to energy $E_{\hbox{\scriptsize av}}$ is defined (with the Boltzmann constant set to 1) by: 
\begin{equation}
 {1 \over T} = \left. {d \ \hbox{log} \ \nu(E) \over d E} \right\vert_{E = E_{\hbox{\scriptsize av}}}.
\label{T}
\end{equation}
The substitution of Gaussian approximation for $\nu(E)$ then gives $T = -\sigma_s/E_{\hbox{\scriptsize av}}$. (Positive temperatures correspond to $E_{\hbox{\scriptsize av}} < 0$.) In a typical situation  of physical interest, $T \sim E_0$, where $E_0$ is a characteristic one-particle energy in the Hamiltonian. Therefore, 
\begin{equation}
 |E_{\hbox{\scriptsize av}}| \sim {\sigma_s \over T}  \gg \sqrt{\sigma_s}.
\label{Eav-typ}
\end{equation}
Like in Eq.(\ref{Emin-sigmas}), the above inequality is the consequence of $N_s \gg \sqrt{N_s}$. Therefore, it obviously extends to all realistic cases of relatively large constituent parts at temperatures in the range $10^{-5} - 10^5$ times $E_0$.

It is shown in Appendix~\ref{spins} and further in Appendix~\ref{condensation} that, under the above conditions, the critical average energy $E_C$ defined by Eq.(\ref{ElEC}) satisfies inequality 
\begin{equation}
 |E_C| \ll \sigma_s \ll |E_{\hbox{\scriptsize min}}|, |E_{\hbox{\scriptsize av}}|.
\label{EC-m}
\end{equation}

Finally, important for proving the condensation into a single lowest level is the property
\begin{equation}
 E_2 - E_{\hbox{\scriptsize min}} \gg {|E_{\hbox{\scriptsize min}}| \over N} \sim {1 \over N \lambda[E_{\hbox{\scriptsize av}}]},
\label{eps2-m}
\end{equation}
where $E_2$ is the energy of the second lowest level.  This inequality is the consequence of the exponential smallness of $N_s$ in comparison to $N$ in combination with the fact that $E_2 - E_{\hbox{\scriptsize min}}$ is, crudely speaking, a single particle property falling on the scale of $|E_{\hbox{\scriptsize min}}|/N_s$ multiplied, perhaps, by some other factors depending polynomially on $N_s$. The rightmost expression in (\ref{eps2-m}) is the consequence of (\ref{lambdaA}).  Inequality (\ref{eps2-m}) is illustrated in Appendix~\ref{spins}.

\subsubsection{Results}

It is shown in Appendix~\ref{condensation} that, when condition (\ref{eps2-m}) is satisfied,  $E_{\lambda}$ is pinned between $E_{\hbox{\scriptsize min}}$ and
$E_2$, sufficiently far from $E_2$, so that 
\begin{equation}
 N [1 + \lambda (E_2 - E_{\hbox{\scriptsize av}})] = 
N {E_2 - E_{\lambda} \over E_{\lambda} - E_{\hbox{\scriptsize av}}}
\gg 1.
\label{lpin}
\end{equation}
This justifies the approximation $p_k \ll 1$ for $k \geq 2$ and, therefore, the validity of formula (\ref{pav2}) for the second lowest level and all levels above it.
In this formula the value of $\lambda$ can then be very accurately approximated by Eq.(\ref{lambdaA}).

As far as the volume $V_1(p_1)$ is concerned, it is narrowly peaked around the maximum, which simultaneously becomes the average value of $p_1$:
\begin{equation}
 \langle p_1 \rangle \approx {E_{\hbox{\scriptsize av}} \over E_{\hbox{\scriptsize min}}}.
\label{p1}
\end{equation}
Such a condensation into the lowest energy state amounts to a significant departure from the result of the conventional microcanonical recipe.

In retrospect, it is also clear that for the case, when the occupation of only one lowest-energy level violates the condition $\langle p_1 \rangle \ll 1$, and therefore formula (\ref{Vkpk1}) does not describe $V_1(p_1)$,  still formula (\ref{pav2}) would give an excellent approximation for all $\langle p_k \rangle$ including $\langle p_1 \rangle$, if the value of $\lambda$ in that formula is found self-consistently from Eq.(\ref{normav}) or (\ref{epsavav}).
The reason is that since formula (\ref{pav2}) is supposed to describe accurately the occupations of all levels beginning from the second, the occupation of the remaining (first) level is bound  by  normalization constraint (\ref{normav}) to have the right value. The self-consistent solution using formula (\ref{pav2}) would produce $E_{\lambda}$ coming as close from below to $E_{\hbox{\scriptsize min}}$ as
necessary in order to reproduce the value (\ref{p1}). As far as other levels are concerned, for the absolute majority of them, the approximation $E_{\lambda} \approx E_{\hbox{\scriptsize min}}$ would remain very accurate independently of whether
$E_{\lambda}$ is slightly above or slightly below $E_{\hbox{\scriptsize min}}$. [Here, a few low-lying levels may constitute a possible exception related to the fact that there is some uncertainty in the present derivation about whether $E_{\lambda}$ stays much closer to $E_{\hbox{\scriptsize min}}$ than to $E_2$. If it does, which I suppose is the case, then the above procedure would be very accurate for all levels with $k \geq 2$.]  

\subsection{The limit of large number of quantum states per particle}
\label{largeS}

When a macroscopic system consists of particles having translational degrees of freedom, the kinetic energy of the particles can reach very high values, before the particles are able to escape from the system. Therefore, the number of quantum states per particle in such a system  can be very large.

In order to analyse this limit in the simplest case, one can consider a system of $N_s$ identical non-interacting oscillators having energy levels equally spaced by $\Omega$ and the average energy per oscillator $n_e \Omega$, where $n_e$ is a finite number.

Each oscillator can, in turn, be described as a large spin $S$  in magnetic field in the limit $S \rightarrow \infty$. The energy of this spin would be 
$E = \Omega S_z$, where the projection $S_z$ admits $(2 S + 1)$ values between $-S$.

In this case, the ground state energy of the whole system is 
\begin{equation}
 E_{\hbox{\scriptsize min}}  = - N_s S \Omega,
\label{EminS}
\end{equation}
while the average energy is 
\begin{equation}
 E_{\hbox{\scriptsize av}}   = - N_s (S - n_e) \Omega.
\label{EavS}
\end{equation}

Therefore, according to formula (\ref{p1}), 
\begin{equation}
 \langle p_1 \rangle \approx {S - n_e \over S} 
\xrightarrow[S\rightarrow \infty]{} 1;
\label{p1S}
\end{equation}
In other words, the most probable state of such a system is the ground state with vanishingly small corrections --- quite a surprising result.

In order to understand it intuitively, one needs to remember that the exact value of $\langle p_1 \rangle$ remains less than one [see Eq.(\ref{p10-result})], and, moreover, $1 -\langle p_1 \rangle \gg 1/N$.  If any eigenstate of the spectrum remains completely unoccupied on average, it means that the corresponding volume in the Hilbert space is zero. Therefore, each of many eigenstates above $E_{\hbox{\scriptsize av}}$ has to have some non-zero average  occupation. At the same time, the eigenstates below $E_{\hbox{\scriptsize av}}$ need to have much greater occupation in order to balance in Eq.(\ref{epsavav}) many more eigenstates above $E_{\hbox{\scriptsize av}}$. It simply turns out that the volume of the Hilbert space is maximized, when almost all (but not all) of the probability weight goes into the ground state.

\subsection{Typical pure state}
\label{typical}

Even though the statistics derived so far has been obtained through averaging over all  possible quantum states subject to the QMC condition, the resulting statistics also describes a typical one among them in the following sense. Once a single  state is selected, it will have very large number of eigenstates in each small energy interval between $E_{\hbox{\scriptsize min}}$ and $E_{\hbox{\scriptsize max}}$. Individually, the occupation numbers of these eigenstates will fluctuate according to the probability distribution proportional to their respective $V_k(p_k)$. That distribution will depend only on the energy of each of these eigenstates, and therefore, within a small energy interval, it will be approximately the same for all of them. Consequently, the average occupation number of eigenstates within any small energy interval will be given by formulas for $\langle p_k \rangle$ obtained above. 

\section{Small subsystem within a large isolated system}
\label{subsystem}

\subsection{Formulation of the problem}
\label{form2}

Now I proceed with deriving the energy distribution for a subsystem of an isolated system --- subject to the QMC condition. It is assumed that the subsystem and the rest of the system --- environment --- do not interact with each other. Therefore, the eigenstates of the whole isolated system can now be labelled by two indices as follows:
\begin{equation}
 \Psi_{\alpha \beta} = \psi_{\alpha} \phi_{\beta},
\label{Psiab}
\end{equation}
where indices $\alpha$ and $\beta$  and the corresponding eigenstates $\psi_{\alpha}$ and $\phi_{\beta}$ refer to the subsystem and the environment respectively. The subsystem has $N_1$ states with energies 
$E_{S \alpha}$ . The environment has $N_2$ states ($N_2 \gg 1$) with energies $E_{E \beta}$. The zero reference point for each set of energies is chosen such that 
\begin{equation}
 \sum_{\alpha=1}^{N_1} E_{S \alpha} = 0,
\label{eSsum}
\end{equation}
and
\begin{equation}
 \sum_{\beta=1}^{N_2} E_{E \beta} = 0.
\label{eEsum}
\end{equation}
The energy of each eigenstate $\Psi_{\alpha \beta}$ of the whole system is then
\begin{equation}
E_{\alpha \beta} =  E_{S \alpha} +  E_{E \beta} .
\label{eab}
\end{equation}
The occupation number of each eigenstate is $p_{\alpha \beta}$.
In this formulation, the density matrix of the subsystem, denoted as 
$\rho_{S \alpha \alpha^{\prime}}$, has only diagonal elements.

I now focus on finding the diagonal element
\begin{equation}
 \rho_{S \alpha \alpha} \equiv \rho_{\alpha} = 
\sum_{\beta = 1}^{N_2} p_{\alpha \beta}.
\label{ra}
\end{equation}
Here I defined variable $\rho_{\alpha}$ just to shorten the notation.

\subsection{General solution}
\label{general}

I now re-label the $N_2$ states contributing to $\rho_{\alpha}$ with index $a$, and the remaining $(N_1 - 1) N_2$ states with index $b$. This results in two new sets of occupation numbers and energies: $(p_a, E_a)$ and $(p_b, E_b)$. Subscripts $a$ and $b$ will play dual role below: as indices and as labels of two different sets. The summation over $a$ implies the first set, and a summation over $b$ implies the second set. In cases, when I refer to the individual members of each set, I use the ``label-and-number'' subscript such as, e.g., $E_{a2}$, which refers to the second lowest energy of the $a$-set, or $E_{b \hbox{\scriptsize min}}$ refers to the minimum energy of the $b$-set.

The new sets of energies have the average values, respectively:
\begin{equation}
 {1 \over N_2} \sum_{a = 1}^{N_2} E_a = E_{S \alpha},
\label{eaav}
\end{equation}
\begin{equation}
 {1 \over N_2 (N_1 - 1)} \sum_{b = 1}^{N_2(N_1 - 1)} E_b = 
- {E_{S \alpha} \over N_1 - 1}
\label{ebav}
\end{equation}

The total normalization constraint and the energy constraints now have form, respectively:
\begin{equation}
 \sum_a p_a + \sum_b p_b = 1,
\label{norm-ab}
\end{equation}
and
\begin{equation}
 \sum_a E_a p_a + \sum_b E_b p_b = E_{\hbox{\scriptsize av}}.
\label{eav-ab}
\end{equation}
Given Eq.(\ref{norm-ab}), condition (\ref{eav-ab}) can be replaced with
\begin{equation}
 \sum_a (E_a - E_{\hbox{\scriptsize av}}) p_a 
+ \sum_b (E_b - E_{\hbox{\scriptsize av}}) p_b 
= 0.
\label{eav-ab0}
\end{equation}

The $a$- and the $b$-states can now divide between themselves the occupations and the total energy as follows:
\begin{equation}
 \sum_a p_a = \rho_{\alpha},
\label{norm-a}
\end{equation}
\begin{equation}
 \sum_a (E_a - E_{\hbox{\scriptsize av}}) p_a  = E_A,
\label{eav-a}
\end{equation}
\begin{equation}
 \sum_b p_b = 1 - \rho_{\alpha},
\label{norm-b}
\end{equation}
\begin{equation}
 \sum_b (E_b - E_{\hbox{\scriptsize av}}) p_b  = - E_A,
\label{eav-b}
\end{equation}
where $E_A$ is the difference between the average energy of the $a$-set and $E_{\hbox{\scriptsize av}}$. It is an auxiliary parameter to be determined simultaneously with $\rho_{\alpha}$.

The goal now is to obtain $\langle \rho_{\alpha} \rangle$ ---   the average value of $\rho_{\alpha}$ over all points in the Hilbert space constrained by conditions (\ref{norm-ab}, \ref{eav-ab0}) in combination with 
\begin{equation}
 p_a, p_b \geq 0.
\label{pa-pb-posit}
\end{equation}

The probability of each pair of values $(\rho_{\alpha}, E_A)$ is  proportional to the volume in the Hilbert space constrained by conditions 
(\ref{norm-a}-\ref{pa-pb-posit}). The constraints on $a$- and $b$-states can then be treated independently by analogy with the problem for the whole system that led to Eq.(\ref{Vkpk2}). This results in the following expression:
\begin{widetext}
\begin{equation}
 V (\rho_{\alpha}, E_A) = V_0 
\hbox{exp}
\left\{
N_2 \hbox{log} \rho_{\alpha} + N_2 (N_1 - 1) \hbox{log} (1- \rho_{\alpha})
+ N_2 \int_{E_{\hbox{\scriptsize av}}}^{E_{\hbox{\scriptsize av}} + {E_A \over \rho_{\alpha}}}
       \lambda_a[E] d E
+ N_2 (N_1 - 1) 
    \int_{E_{\hbox{\scriptsize av}}}^{E_{\hbox{\scriptsize av}} - {E_A \over 1-\rho_{\alpha}}}
       \lambda_b[E] d E
\right\},
\label{V-rho}
\end{equation}
\end{widetext}
where $\lambda_a(E)$ and $\lambda_b(E)$ are the parameters analogous to $\lambda$ introduced below for the spectrum of the entire system, but this time defined for the spectra of $E_a$ and $E_b$ separately, and $V_0$ is an unimportant prefactor.

Due to the fact that $N_2 \gg 1$, the expression (\ref{V-rho}) must be very sharply peaked near the maximum of the function in the power of the exponent. Therefore finding $\langle \rho_{\alpha} \rangle$ is reduced to finding the value of $\rho_{\alpha}$ at the maximum of this power. In order to locate that maximum, I look for the zeros of the partial derivatives of the exponential power in (\ref{V-rho}) with respect to $E_A$ and $\rho_{\alpha}$. Differention with respect to $E_A$ gives
\begin{equation}
 {N_1 - 1 \over 1- \rho_{\alpha}}  \ 
\lambda_b\left[E_{\hbox{\scriptsize av}}- {E_A \over 1-\rho_{\alpha}}\right] 
= {
\lambda_a\left[E_{\hbox{\scriptsize av}}+ {E_A \over \rho_{\alpha}}\right]
\over 
\rho_{\alpha}
}.
\label{cond1}
\end{equation}
Now, differentiating the power in Eq.(\ref{V-rho}) with respect to $\rho_{\alpha}$ and also using (\ref{cond1}), I obtain
\begin{equation}
 \rho_{\alpha} ( 1 - N_1 \rho_{\alpha}) 
= E_A \ 
\lambda_a\left[E_{\hbox{\scriptsize av}}+ {E_A \over \rho_{\alpha}}\right].
\label{cond2}
\end{equation}

One can get a useful insight into the solution of Eqs.(\ref{cond1},\ref{cond2}) by substituting $\lambda_a$,$\lambda_b$ with  new variables $E_{\lambda_a}$, $E_{\lambda_b}$:  
\begin{equation}
 \lambda_a [E] = {1 \over E - E_{\lambda_a}[E] }
\label{lambda-a}
\end{equation}
\begin{equation}
 \lambda_b [E] = {1 \over E - E_{\lambda_b}[E] }
\label{lambda-b}
\end{equation}
Subscripts $a$ and $b$ in variables $\lambda_a$,$\lambda_b$, $E_{\lambda_a}$, $E_{\lambda_b}$ are just the labels of the characteristics of the respective energy spectra, i.e. they are not indices running over a set of integer values.

After some manipulations, the above substitution generates two equations equivalent to (\ref{cond1},\ref{cond2}):
\begin{equation}
 E_A = \left( {1 \over N_1} - \rho_{\alpha} \right) 
\left\{
E_{\hbox{\scriptsize av}} - E_{\lambda_a}\left[E_{\hbox{\scriptsize av}}+ {E_A \over \rho_{\alpha}}\right]
\right\}
\label{EA}
\end{equation}
\begin{equation}
 E_{\lambda_a}\left[E_{\hbox{\scriptsize av}}+ {E_A \over \rho_{\alpha}}\right]
=
E_{\lambda_b}\left[E_{\hbox{\scriptsize av}}- {E_A \over 1-\rho_{\alpha}}\right].
\label{Ela-Elb}
\end{equation}
The latter equation is the key to the following solution for the macroscopic environment.

In general, Eqs.(\ref{cond1}, \ref{cond2}), or equivalently Eqs.(\ref{EA},\ref{Ela-Elb}) should be solved numerically.
However, for the case of a small subsystem and a macroscopic environment, and with the realistic value of $E_{\hbox{\scriptsize av}}$ for the whole system (as discussed in Section~\ref{macro}), the system of Eqs.(\ref{EA},\ref{Ela-Elb}) can be solved analytically. 

There is also another analytically solvable limit, which corresponds to the case of high Hilbert space temperatures, i.e. very small $\lambda_a$ and $\lambda_b$. This limit is not to be considered in this work.

\subsection{Subsystem in a macroscopic environment}
\label{macro-env}

The condition of macroscopic environment and a small subsystem amounts formally to the presence of narrow Gaussian-like maximum around $E = 0$ in the density of states of the environment $\nu_E(E)$ with the mean-squared spread of energies $\sigma_E$ satisfying the conditions:
\begin{equation}
 |E_{E \hbox{\scriptsize min}}|, E_{E \hbox{\scriptsize max}} , |E_{\hbox{\scriptsize av}}|  \gg \sqrt{\sigma_E} \gg  |E_{S \alpha}|, \ \  \forall \alpha.
\label{s2A}
\end{equation}
In addition, there is a reasonable condition for the differences between two lowest energy states for the subsystem and the environment:
\begin{equation}
 E_{S2} - E_{S \hbox{\scriptsize min}} \geq E_{E2} - E_{E \hbox{\scriptsize min}}.
\label{l-space}
\end{equation}

It is shown in Appendix~\ref{subsystem-App} that, in this case, the occupations of the lowest subsystem state in the leading order of $E_{S \alpha} / E_{E \hbox{\scriptsize min}}$ is 
\begin{equation}
 \langle \rho_1 \rangle  = 
{ E_{\hbox{\scriptsize av}}  \over  E_{E \hbox{\scriptsize min}} } 
+
{1 \over N_1} \
\left( 1 - 
{ E_{\hbox{\scriptsize av}}  \over  E_{E \hbox{\scriptsize min}} } 
\right)
\label{rho1-av}
\end{equation}
and, for the remaining states with $\alpha \geq 2$,
\begin{equation}
 \langle \rho_{\alpha} \rangle  = 
{1 \over N_1} \
\left( 1 - 
{ E_{\hbox{\scriptsize av}}  \over  E_{E \hbox{\scriptsize min}} } 
\right),
\label{rhoa-av}
\end{equation}
where the right-hand side is obviously independent of $\alpha$.

It is also possible to obtain more general formulas (see Appendix~\ref{subsystem-App}):
\begin{equation}
 \langle \rho_1 \rangle  = 
{
E_{\hbox{\scriptsize av}} (1 - {1 \over N_1}) + {E_{S \hbox{\scriptsize min}} \over N_1 -1} 
+ {E_{\hbox{\scriptsize min}} \over N_1} 
\over 
E_{\hbox{\scriptsize min}} + {E_{S \hbox{\scriptsize min}} \over N_1 -1}
},
\label{rho1-avA}
\end{equation}
and, for $\alpha \geq 2$,
\begin{equation}
 \langle \rho_{\alpha} \rangle  = {1 \over N_1} {E_{\hbox{\scriptsize av}} - E_{E \hbox{\scriptsize min}}  \over E_{S \alpha} -  E_{E \hbox{\scriptsize min}} },
\label{rhoa-avA}
\end{equation}
where $E_{\hbox{\scriptsize min}} = E_{S \hbox{\scriptsize min}} + E_{E \hbox{\scriptsize min}}$.
Formulas (\ref{rho1-avA},\ref{rhoa-avA})  are certainly valid up to the first order in $E_{S \alpha} / E_{E \hbox{\scriptsize min}}$, but in fact have a broader range of applicability, because conditions (\ref{s2A},\ref{l-space}) are sufficient but not necessary for the validity of approximation (\ref{rho1-avA},\ref{rhoa-avA}). For example, this approximation also describes the case $E_{S \alpha} \sim E_{E \hbox{\scriptsize min}}$, when condition $|E_{S \alpha}| \ll |E_{E \hbox{\scriptsize min}}|$ is replaced by the requirement that $N_1 \gg 1$ and the density of states for energies $E_{S \alpha}$ have a Gaussian-like narrowly peaked shape around $E = 0$. Further discussion of the necessary conditions for the validity of approximation (\ref{rho1-avA},\ref{rhoa-avA}) is given in Appendix~\ref{subsystem-App}.

The basic assumptions leading to the above results contain  a loophole of neglecting the interaction between the subsystem and the environment. The same loophole is also present in the conventional micro-canonical derivation of the Boltzmann-Gibbs statistics. In principle, given the condensation of the entire system into the lowest energy state, one should not be surprised that a similar property is exhibited by a subsystem. Yet one can still worry about the validity of the condensation into the single lowest energy state of the subsystem [Eq.(\ref{rho1-av})], when the interaction energy with the environment is much greater than the separation between the lowest and the second lowest energy levels of the subsystem.
In this case, the occupation numbers of the subsystem would depend on entanglement properties with the environment in the ground state of the whole system. This loophole potentially opens the window for chaos and non-integrability to play a role in the resulting statistics. It also cannot be excluded that the result may then reproduce the Boltzmann-Gibbs statistics in the   energy range of the order of the subsystem-environment interaction energy.
This issue is not addressed further in the present work.

\subsection{Typicality for the pure states of the whole system for the density matrix of a subsystem}
\label{typicality}

The probability distribution of parameters describing the density matrix of a small subsystem of the whole system exhibits exponentially narrow ($\sim 1/N_2$) maxima controlling the average values of these parameters. Therefore, a random choice of QMC-constrained single quantum state would be, with probability $1- O(1/N_2)$, exponentially close to the average values computed above. This situation is analogous to the ``canonical typicality''\cite{Popescu-06,Goldstein-06} for the conventional microcanonical condition. 

\section{Numerical tests}
\label{numerical}

The analytical results of this work  and there relevance to realistic systems should be checked numerically. Here, I present two such tests of preliminary nature
addressing only the statistics for the entire isolated quantum system. 

\subsection{Direct random sampling}
\label{direct}


\begin{figure} \setlength{\unitlength}{0.1cm}

\begin{picture}(100, 130)
{ 
\put(5,125){\textsf{\Large (a)}}
\put(0, 65){ \epsfxsize= 3.3in \epsfbox{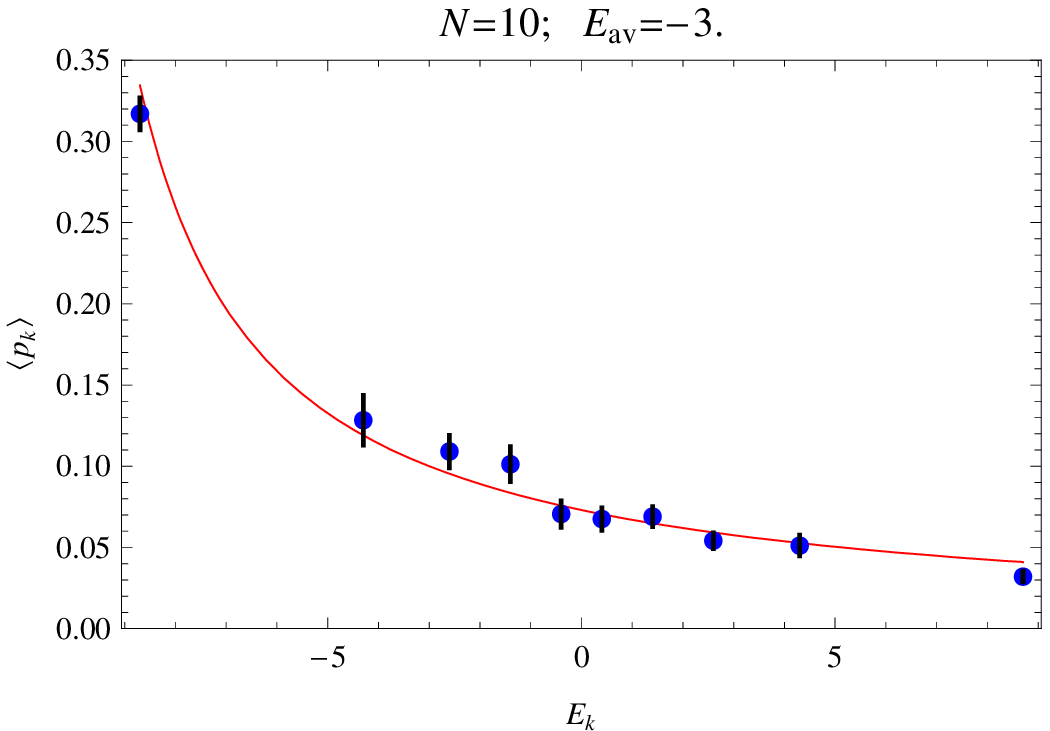} }
\put(5,60){\textsf{\Large (b)}}
\put(2, 0){ \epsfxsize= 3.35in \epsfbox{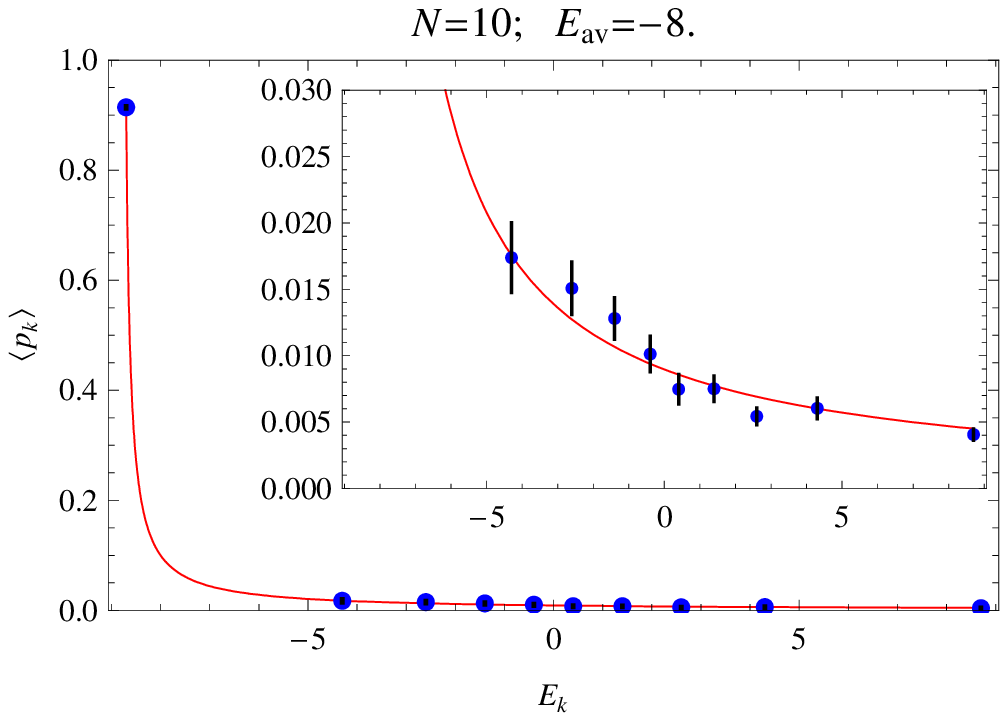} }
}
\end{picture} 
\caption{(Color online) Two results of direct random sampling of 
quantum states in a system of ten energy levels. Average energies are indicated above the plots. Dots with error bars represent the averages over all sampled states. Solid lines represents the prediction of Eq.(\ref{pav2}) with the value of $\lambda$ obtained numerically from formula Eq.(\ref{epsavav}). The inset in plot (b) magnifies the small $\langle p_k \rangle$ part of the main plot.
} 
\label{fig-direct} 
\end{figure}


The first test is a direct Monte-Carlo sampling of the Hilbert space under constraint (\ref{epsav}). I have done this sampling directly in the Eucledean space of variables $\{ p_i \}$ using the algorithm consisting of the following steps: (i) selection of an orthonormal basis in the $(N-2)$-dimensional hyperplane constrained by Eqs.(\ref{epsav}, \ref{norm}); (ii) identification in that hyperplane a $(N-2)$-dimensional hypercube, which encloses all the vertices of the intersection manifold; (iii) random sampling of points within that hypercube; and, finally (iv) acceptance of only those random points, which in the original $N$-dimensional $\{ p_i \}$-space have all non-negative coordinates as required by constraint (\ref{positiv}).

This algorithm is not very efficient: its acceptance rate at step (iv) decreases by about factor of 10 as $N$ increases by one. Using {\it Mathematica} software, I was able to generate statistically significant number of random points for the case of $N=10$ with the specrum and the average energy shown in Fig.~\ref{fig-direct}.

This figure  compares the average occupation numbers obtained numerically with the approximate theoretical values obtained on the basis of Eq.(\ref{pav2}) with with the value of $\lambda$ found by solving Eq.(\ref{epsavav}) numerically.

The lowest level in Fig.~\ref{fig-direct}(b)  violates the condition $\langle p_k \rangle \sim 1/N \ll 1$. However, as discussed at the end of Section~\ref{macro}, the overall structure of the more accurate  solution guarantees that, for a single level violating the above condition, the result (\ref{epsavav}) would still amount to a very good approximation, even though the corresponding Hilbert space volume $V_1(p_1)$ is does not any longer decay exponentially but instead is peaked around $p_1 \approx \langle p_1 \rangle$.

Given that $N=10$ barely qualifies as a very large number, the overall agreement exhibited in Fig.~\ref{fig-direct} is surprisingly good.


\begin{figure} \setlength{\unitlength}{0.1cm}

\begin{picture}(100, 195)
{ 
\put(5,190){\textsf{\Large (a)}}
\put(0, 130){ \epsfxsize= 3.3in \epsfbox{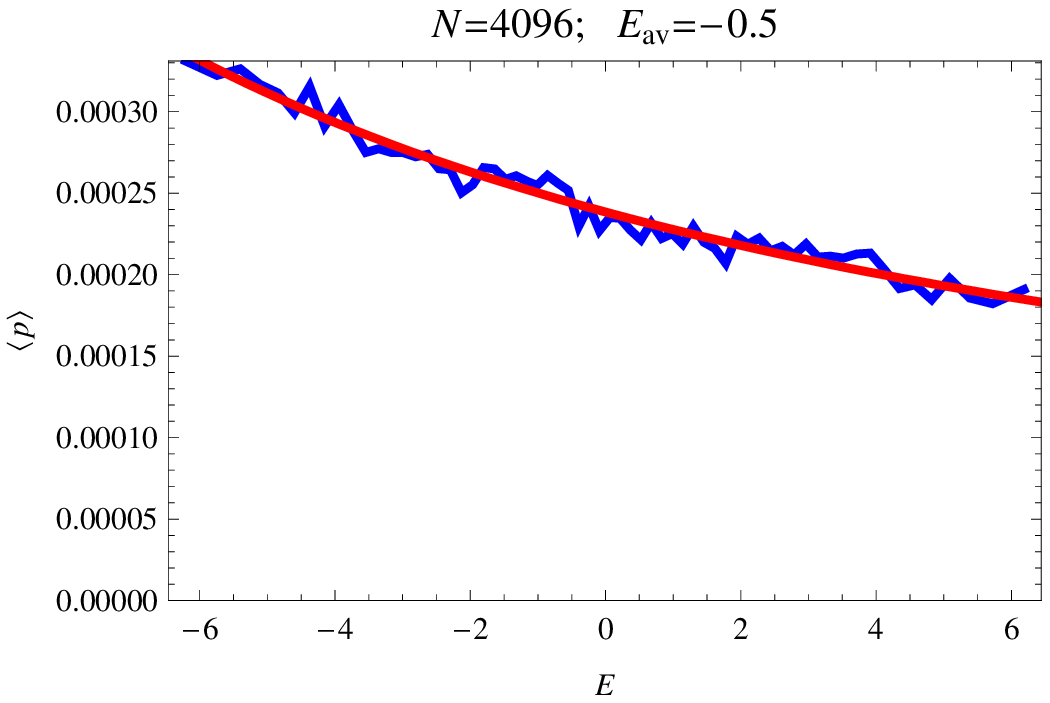} }
\put(5,125){\textsf{\Large (b)}}
\put(0, 65){ \epsfxsize= 3.3in \epsfbox{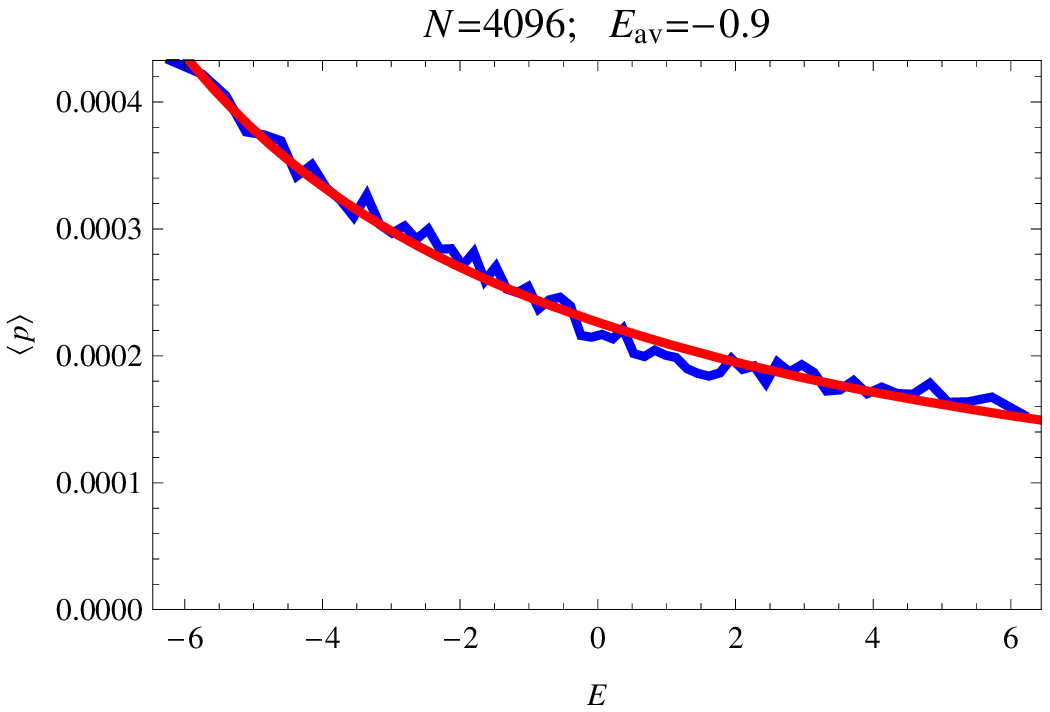} }
\put(5,60){\textsf{\Large (c)}}
\put(0, 0){ \epsfxsize= 3.3in \epsfbox{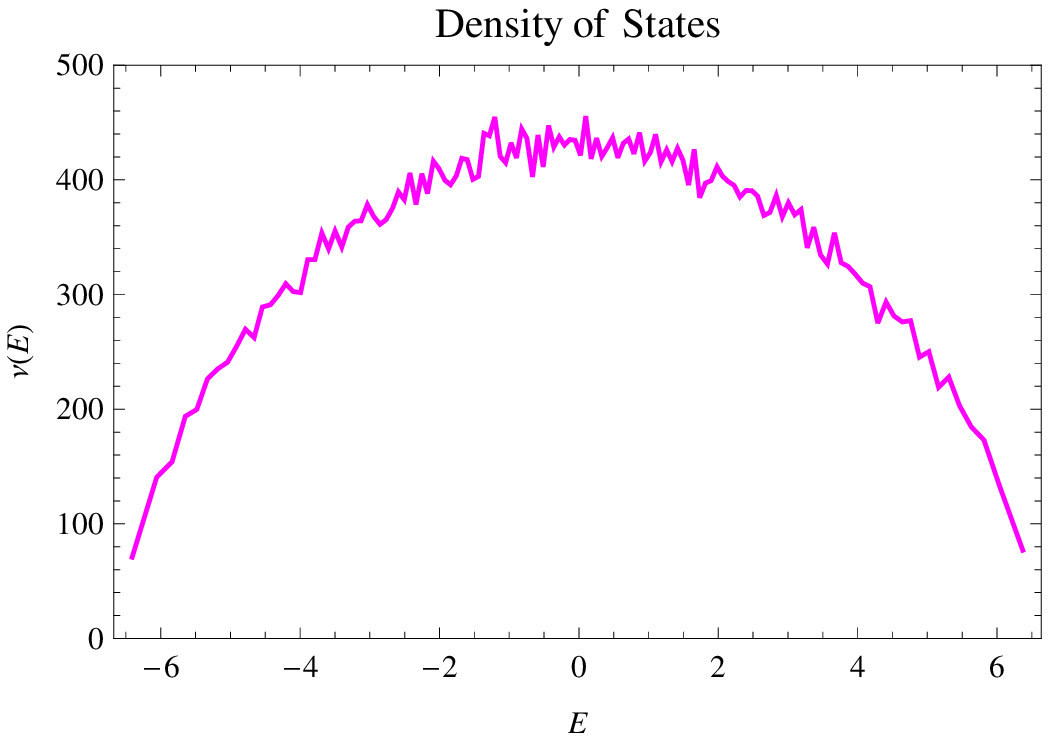} }
}
\end{picture} 
\caption{(Color online) 
Average occupations of eigenstates $\langle p \rangle$ participating in the expansions of non-eigenstates of the initial basis, which (the non-eigenstates) are selected from a narrow energy window $E_{\hbox{\scriptsize av}} \pm 0.01$ around the values indicated above plots (a) and (b). Broken lines in these two plots represent numerical results  averaged over groups of 64 adjacent eigenstates. Smooth lines represents the prediction of Eq.(\ref{pav2}) with the values of $\lambda$ obtained numerically from Eq.(\ref{epsavav}). The density of eigenstates $\nu(E)$ for this system averaged over groups of 32 states is presented in plot (c). 
} 
\label{fig-RandMatr} 
\end{figure}


\subsection{Random matrices}
\label{random}

Now I show that the statistics obtained in this work manifests itself in a system described by a random matrix Hamiltonian.  

If one of the original non-eigenstates in the basis, where the random matrix is defined, has energy $E_{\hbox{\scriptsize av}}$ (the diagonal element of the random matrix corresponding to that state), this imposes constraint (\ref{epsav}) on the eigenstates participating in the expansion of the selected state. The present numerical experiment was based on a guess, successfully confirmed by the end result, that the eigenstates in this case would participate in the expansion of a non-eigenstate, as if that expansion was done randomly on the basis of the QMC condition.

I took a $4096 \times 4096$ matrix, where all diagonal elements and a fraction ${30 \over 4096}$  of off-diagonal elements were assigned random values picked in the interval $[-1,1]$. The remaining off-diagonal elements were zeros. The Hamiltonian was diagonalized, and then one state of the original non-eigenbasis was chosen and expanded in the eigenstate basis.  The weight of individual eigenstates fluctuated as expected from Eq.(\ref{Vkpk1}). However, once the spectrum is divided in groups of 64 adjacent eigenstates having approximately the same energy,  then the average weight within each group begins converging to the theoretical approximation  (\ref{pav2}) as discussed in Section~\ref{typical}.  I further improve the error bars by combining the overall statistics for $34-35$ non-eigenstates with average energies within a narrow energy window $E_{\hbox{\scriptsize av}} \pm 0.01$, where  $E_{\hbox{\scriptsize av}}$ is equal to $-0.5$ and $-0.9$ in the two examples shown in Fig.~\ref{fig-RandMatr}. The good agreement is then revealed with the theoretical approximation (\ref{pav2}), which uses $\lambda$ computed numerically from Eq.(\ref{epsavav}).

When I increase the fraction of non-zero off-diagonal elements in the random matrix, the agreement between numerics and the theory continues to hold. In this case, however, the width of the nearly semi-circle eigenspectrum increases, while the window for $E_{\hbox{\scriptsize av}}$ determined by the diagonal elements of the Hamiltonian remains the same.  As a result, $\lambda$ becomes small and the dependence (\ref{pav2}) becomes difficult to distinguish from a linear one. On the other hand, if the fraction of non-zero off-diagonal elements decreases, then the assumption of perfect mixing of eigenstates in each of the original basis states becomes increasingly inadequate, as the weights of eigenstates start to peak around $E = E_{\hbox{\scriptsize av}}$. Such a behavior is natural to expect as this system gradually approaches the limit of small off-diagonal elements in the Hamiltonian (see, e.g., Ref.\cite{Tasaki-98}).

\section{Concluding remarks}
\label{concluding}

1) The results presented in this work, indicate that the statistical description of an isolated quantum system subject to a fixed energy constraint and unrestricted participation of eigenstates contradicts (at least in the limits considered) 
to the Boltzmann-Gibbs statistics derivable on the basis of the conventional microcanonical assumption.

In particular, the resulting energy distributions for both the whole isolated system [Eq.(\ref{pav2})] and a small subsystem of it [Eq.(\ref{rhoa-av})]  show algebraic rather than exponential dependence on the energies of participating states as well as routine macroscopic occupancy (condensation) for the lowest-lying energy states --- Eqs.(\ref{p1},\ref{p1S},\ref{rho1-av}).

One should be mindful though of the loophole associated with the neglected interaction between the subsystem and the environment --- see the discussion at the end of Section~\ref{macro-env}.

2) The statistics derived in this work is supported by the numerical findings presented in Section~\ref{numerical}. Particularly interesting is the finding presented in Section~\ref{random},
that the expansion of the non-eigenstates of the random matrix basis in terms of eigenstates follows this statistics for large but finite random matrices, which are not too sparce.

3) The finite size version this statistics might thus be observable after a strong (and preferably non-integrable) perturbation of a well isolated system having a relatively small number of particles but a relatively large number of quantum levels. Nanoscale-limited systems should be good candidates for such a study. For a better control of the total energy after the perturbation, one can proceed in analogy with the ``numerical experiment'' on a random matrix presented in Section~\ref{random}. Namely, one can force the system into a single quantum state, e.g. the ground state, before perturbing it.

4) In systems of bosons, the condensation described in this work into the ground or a few lowest states may produce an appearance of Bose-Einstein condensation, because the ground state is indeed Bose-condensed. Yet the nature of the two kinds of condensations is different. The former represents a jump in the occupation of many-particle states, while the latter is a single-particle phenomenon accompanied by the usual exponential statistics for the occupations of all many-particle quantum states.

5)  When it comes to macroscopic systems, the contradiction indicated in remark 1 reenforces the concern that non-relativistic quantum mechanics alone is not sufficient to justify the Boltzmann-Gibbs statistics. One needs an assumption of external origin, such as the quantum collapse of the broad distribution of eigenstates into a narrow energy window postulated by the conventional microcanonical description.

If such a collapse happens even once, it appears very difficult if not impossible to realistically perturb a typical macroscopic system containing many weakly interacting parts into a state characterized by a broad energy range of participating eigenstates.

Yet, if this collapse occurs continuously, its description would go beyond the linear quantum mechanics and in particular may imply additional source of energy fluctuations for the entire system.

6) Even though the statistics obtained in this work appears to contradict to the everyday experience well describable by the Boltzmann-Gibbs statistics, it is still interesting to think, what the present statistics might imply, if one assumes that the entire Universe is describable by a single wave function.

7) Quite a few researchers including this author (see e.g. Refs.\cite{Deutsch-91,Srednicki-94,Gaspard-00,Rigol-08,Fine-04-chaos,Fine-05-echo,Morgan-08}) share the feeling that the elusive notion of quantum chaos plays an important role in the foundations of quantum statistical physics. Chaos, however, plays no role in finding the most probable quantum state both in the present  and in the conventional micro-canonical formulation. This suggests that the role of chaos is not to determine the equilibrium itself but rather to influence how a subsystem  relaxes towards the equilibrium.

Yet, one should not forget about the loophole related to the neglected interaction between the subsystem and environment [see the end of Section~\ref{macro-env}].  In addition, what quantum chaos certainly does is that it strongly suppresses the fluctuations of the number of energy levels within any fixed energy window --- consequence of the repulsion of energy levels. Whether and how this property would affect the macroscopic characterization of equilibrium in quantum systems is not clear to this author.

8) From a broader perspective,  the statistics  based on simple constraints (\ref{epsav}, \ref{norm}, \ref{positiv}) describes a distribution of an essentially positive and limited in the amount quantity $p$ among $N$ agents having characteristics $\{ E_k \}$. 
It is, therefore, tempting to speculate that such a statistics  might be applicable beyond the quantum mechanical problems, in particular, to the problems of economics, when and if one finds a meaningful interpretation for constraint (\ref{epsav}).

\acknowledgements

The author is grateful to D. Antonov, P. Gaspard, T. Gr\"{u}newald, A. Komnik, F. Wegner for discussions related to the content of this work and to W. Wustmann for early computer studies. A significant part of this work was done during author's stay at the University of Tennessee, Knoxville.

\appendix

\section{Proof that all values of $\lambda_k$ are approximately equal}
\label{central}

In order to be specific, let us compare the intersection manifolds $M_1$ and $M_2$ pertaining respectively to $\lambda_1$ in the space of all variables $p_i$ excluding $p_1$ and $\lambda_2$ in the space of all variables $p_i$ excluding $p_2$. Let us also, for this part only, shift the origin of the energy zero point to $E_{\hbox{\scriptsize av}}$, i.e. the condition $E_{\hbox{\scriptsize av}}=0$ replaces Eq.(\ref{epsref}); and assume that $E_1$ and $E_2$ are not necessarily two lowest energies of the spectrum, but rather two arbitrary ones.

In the first case, the intersection manifold is defined by
\begin{equation}
p_2, p_3, ..., p_N \geq 0,
\label{positM1}
\end{equation}
\begin{equation}
p_2 + p_3 + ... + p_N = 1,
\label{normM1}
\end{equation}
\begin{equation}
E_2 p_2 + E_3 p_3 + ... + E_N p_N = v,
\label{epsavM1}
\end{equation}
where $v= - E_1 p_1$.

Parameter $\lambda_1$ characterizes the change of volume $V_1$ 
of the above manifold in response to small shift $v$ of hyperplane (\ref{epsavM1}). Specifically,
\begin{equation}
\lambda_1 =  \left. {1 \over V_1} {d V_1 \over dv} \right\vert_{v=0} =
\sum_{\alpha = 1}^{N-3} {1 \over \eta_{1 \alpha}} {d\eta_{1 \alpha} \over dv} =
\sum_{\alpha = 1}^{N-3} \lambda_{1 \alpha}.
\label{lambda1}
\end{equation}

In the second case, the intersection manifold is 
\begin{equation}
p_1, p_3, ..., p_N \geq 0,
\label{positM2}
\end{equation}
\begin{equation}
p_1 + p_3 + ... + p_N = 1,
\label{normM2}
\end{equation}
\begin{equation}
E_1 p_1 + E_3 p_3 + ... + E_N p_N = v,
\label{epsavM2}
\end{equation}
where $v= - E_2 p_2$.

Parameter $\lambda_2$ characterizes the change of volume $V_2$ 
of the above manifold:
\begin{equation}
\lambda_2 =  \left. {1 \over V_2} {d V_2 \over dv} \right\vert_{v=0} =
\sum_{\alpha = 1}^{N-3} {1 \over \eta_{2 \alpha}} {d\eta_{2 \alpha} \over dv} =
\sum_{\alpha = 1}^{N-3} \lambda_{2 \alpha}.
\label{lambda2}
\end{equation}

If one changes the variable $p_1$ to $p_2$ in Eqs.(\ref{positM2}-\ref{epsavM2}), then one finds that the problems of calculating $\lambda_1$ and $\lambda_2$ are nearly identical to each other with the only difference that term $E_2 p_2$ in Eq.(\ref{epsavM1}) has to be replaced with $E_1 p_2$. One can already anticipate the result $\lambda_1 \approx \lambda_2$ from the fact that vectors
$\{ E_2, E_3, ..., E_N \}$ and 
$\{ E_1, E_3, ..., E_N \}$ determining the normal directions of the hyperplanes (\ref{epsavM1}) and (\ref{epsavM2}), respectively,
are nearly parallel to each other --- consequence of the large number of identical components. the proof given below, however, takes a different route.

It is, in fact, possible to show, that the sets of characteristic linear parameters $\{ \eta_{1 \alpha} \}$ and $\{ \eta_{2 \alpha} \}$ can be chosen such that $ \eta_{1 \alpha}(v) = \eta_{2 \alpha} (v) $ for all but one value of $\alpha$, and, therefore, $N-4$ out of $N-3$  terms contributing to $\lambda_1$ and $\lambda_2$ in Eqs.(\ref{lambda1}, \ref{lambda2}) are identical to each other. Therefore, in general,
\begin{equation}
 \lambda_2 = \lambda_1 + O(1/N).
\label{l1l2}
\end{equation}

The proof of the above statement lies in the fact that two $(N-3)$-dimensional manifolds $M_1$ and $M_2$ (with the $p_1$-to-$p_2$ variable change in the second case)  have a common $(N-4)$-dimensional face. This face is defined by condition $p_2 = 0$. Therefore, one can choose the first linear parameter $\eta$ in each case differently by dividing the volumes $V_1$ and $V_2$ by the $(N-4)$-dimensional volume of that face, but then the rest of the parameters can be chosen identically, because they will describe the volume renormalization for the same face.  

I now describe the above $(N-4)$-dimensional face in more detail and, in particular show, that the $(N-1)$-dimensional hyperplane $p_2 = 0$ contains more than $N-3$ vertices of $M_1$ or $M_2$ --- a necessary condition to form a $(N-4)$-dimensional face.

All the vertices of either manifold $M_1$ or $M_2$ are obtained by the intersection of the linear edges of the normalization Hypertetrahedron (defined in Section~\ref{pk-less-1}) with the energy hyperplane. In turn, all of these edges have form 
$p_i = 1-p_j$, $(0 \leq p_i, p_j \leq 1)$  in the two-dimensional plane defined by condition $p_k = 0,  \forall k\neq i,j$.

The energy hyperplane (with $v=0$) intersects such an edge only in the case, when $E_i$ and $E_j$ have opposite signs. (Here and everywhere below I ignore the non-generic case of $E_i$ or $E_j$ equal to zero.) Therefore, if there are $K$ quantum states with $E_i > 0$ [$E_i > E_{\hbox{\scriptsize av}}$] and $L = N-1-K$ with $E_i < 0$ [$E_i < E_{\hbox{\scriptsize av}}$], then the intersection manifold has $KL$ vertices.

Among these $KL$ vertices, only those originating from the edges involving $p_2$
change between the manifolds $M_1$ and $M_2$ (shift or disappear):

if $E_1, E_2 > 0$, then $L$ vertices shift, possibly significantly;

if $E_1, E_2 < 0$, then $K$ vertices shift;

if $E_1$ and $E_2$ have opposite signs, then $L$ vertices in one case are replaced by $K$ vertices  in the other case.

All other vertices remain identical. They all lie in the hyperplane $p_2 =0$.
Their number is greater than $KL- (N-1)$, which, in turn is greater (normally, much greater) than $(N-3)$, when $K,L > 2$. Therefore  this number  is sufficient to form an $(N-4)$-dimensional face.

One can further show that the above vertices do not fall into a lower dimensional subspace but instead, can be used to form $(N-4)$ linearly independent vectors. The coordinates of these vectors can have $(N-2)$ projections in the original Hilbert space: $p_3, p_4, ..., p_N$. Let us further assume that $E_3 < 0$ and $E_4 > 0$. Therefore, there will be one vertex  in the $\{p_3,p_4\}$ two-dimensional plane. I denote this vertex as $w_{34}$. In addition, there will be $K^{\prime}$ vertices in $\{p_3,p_i\}$-planes, such that $i \geq 5$ and $E_i >0$; and
$L^{\prime}$ vertices in $\{p_4,p_j\}$-planes, such that $j \geq 5$ and $E_j < 0$. The total number of vertices in the latter two sets is $K^{\prime} + L^{\prime} = N-4$. One can now form $(N-4)$ vectors by subtracting the coordinates of vertex $w_{34}$ from each of the above $(N-4)$ vertices. The resulting vectors will be linearly independent, because the projections on each of the axes $p_5, p_6, ..., p_N$ will be present in this set only once in one of the vectors, and therefore the linear combination with other vectors cannot cancel that projection. 

Finally, the above $(N-4)$ dimensional face changes identically under the shift of the energy hyperplanes in the both cases (which includes the possible change in the number of vertices). Indeed, because of the condition $p_2=0$, the projected shifts of the energy hyperplanes in both cases are described by identical equation:
\begin{equation}
E_3 p_3 + ... + E_N p_N = v.
\label{epsavMv}
\end{equation}
Therefore, all $(N-4)$ renormalization coefficients describing the volume change of this face with $v$ can be chosen identically. 

As indicated in the discussion of manifold vertices, the structures of the two manifolds compared can be very different outside the common face. However, all these differences contribute to a single linear parameter $\eta$ corresponding to the direction perpendicular to the common face, which changes only weakly under the change of $v$.

The last step for the complete rigor of the present proof would be to impose the limits on the exceptional cases, when the renormalization of a single parameter $\lambda_{k \alpha_0}$ would be comparable with the sum of the rest of $\lambda_{k \alpha}$. 

One possibility, of course, is that  the latter sum accidentally turns to zero,
even though each term of that sum is comparable to $\lambda_{k \alpha_0}$. However, this would actually mean that the typical $\lambda$, as used in the main text, is zero within the accuracy of the present approximation.

The real problematic case would correspond to $\lambda_{k \alpha_0}$ being much greater than almost all of other $\lambda_{k \alpha}$. Such a situation appears to arise only in the case of little physical interest, namely, when $|E_k|$  is of the order of the root-mean-squared value of all other energies, but this claim requires further proof. Here, I would only like to comment that the common face $(N-4)$-dimensional face is highly unlikely to be uncharacteristic of the $(N-3)$-dimensional manifold, because, if all the vertices of this face are removed from the manifold, the remaining ones will not be enough to form an $(N-3)$-dimensional manifold (when $K>1$ and $L>1$ ).

\section{Derivation of general formula (\ref{Vkpk2}) for the Hilbert space volume of manifold $M_k$}
\label{integral}

The derivation of formula (\ref{Vkpk2}) proceeds as follows.

Both sides in each of Eqs.(\ref{normK}, \ref{epsavK0}) are divided by $(1-p_k)$ and new variables
\begin{equation}
 p_i^{\prime} = {p_i \over 1 - p_k}
\label{p-prime}
\end{equation}
for $i \neq k$ are introduced. This gives, respectively,
\begin{equation}
\sum_{i, i\neq k}^N p_i^{\prime} = 1,
\label{normK-pr}
\end{equation}
\begin{equation}
\sum_{i, i\neq k}^{N}  (E_i - E_{\hbox{\scriptsize av}})  p_i^{\prime}   = v,
\label{epsavK0-pr}
\end{equation}
where
\begin{equation}
 v = - {(E_k - E_{\hbox{\scriptsize av}}) p_k \over 1- p_k}
\label{v}
\end{equation}
not a small number.
The above equations in combination with constraints
\begin{equation}
 p_i^{\prime} \geq 0
\label{posit-prime}
\end{equation}
describe a manifold $M_k^{\prime}$ with volume $V_k^{\prime}$ in the space of variables $p_i^{\prime}$. Due to rescaling of axes (\ref{p-prime}),
\begin{equation}
 V_k = (1-p_k)^{N-3} \  V_k^{\prime}
\label{V-Vpr}
\end{equation}
where $V_k$ is the volume of the original manifold $M_k$ defined in Section~\ref{form1}, and power $N-3$ is equal to the dimension of manifolds 
$M_k^{\prime}$ and $M_k$.

Since the set of energy coefficients in the left-hand side of Eq.(\ref{epsavK0-pr}) for manifold $M_k^{\prime}$ is the same as the set for the original manifold $M_k$ in Eq.(\ref{epsavK0}), both manifolds are characterized by the identical dependence of the parameter $\lambda$ on the shift of the energy hyperplanes:
\begin{equation}
 \lambda[E_{\hbox{\scriptsize av}}] = {1 \over N-3} \ {d \ \hbox{log} V^{\prime} \over d v} = 
 {1 \over N-3} \ {d \ \hbox{log} V^{\prime} \over d E_{\hbox{\scriptsize av}}}
\label{lambda-der}
\end{equation}
(see also Appendix~\ref{central}). Therefore, 
\begin{equation}
 V^{\prime} = V_0^{\prime} \hbox{exp}
\left\{
(N-3)
\int_{E_{\hbox{\scriptsize av}}}^{E_{\hbox{\scriptsize av}} + v} \lambda[E] dE
\right\},
\label{Vprime}
\end{equation}
where $V_0^{\prime}$ is the volume of $M_k^{\prime}$ corresponding to $v=0$,
which is simultaneously equal to $V_k$, when $p_k=0$. Combining Eqs.(\ref{V-Vpr}), (\ref{Vprime}) and (\ref{v}), I obtain
\begin{widetext}
\begin{equation}
 V_k(p_k) = V_k(0) \hbox{exp} \left\{
(N-3)
[ \hbox{log}(1-p_k) + \int_{E_{\hbox{\scriptsize av}}}^{E_{\hbox{\scriptsize av}} - {(E_k -E_{\hbox{\scriptsize av}}) p_k \over 1-p_k} } \lambda[E] d E
\right\},
\label{Vkpk2-App}
\end{equation}
\end{widetext}
the same as Eq.(\ref{Vkpk2}) in the main text. This equation has to be solved together with Eq.(\ref{epsavav}), which requires finding $\lambda[E]$ for  most of the allowed interval $[E_{\hbox{\scriptsize min}}, E_{\hbox{\scriptsize max}}]$. In practice, however,
if the purpose is to find $V_k(p_k)$ for a particular value of $E_{\hbox{\scriptsize av}} <0$, the interval $[E_{\hbox{\scriptsize av}}, 0]$ would surfice. This self-consistent solution can proceed as follows:

One first finds the values of  $\lambda[E_{\hbox{\scriptsize av}}]$ in the interval, where the approximation (\ref{pav2}) is valid, i.e. for $E_{\hbox{\scriptsize av}}$ between $0$ and a certain value somewhat below the critical value $E_C$ defined by Eq.(\ref{ElEC}) , and then proceed with reducing $E_{\hbox{\scriptsize av}}$ further in sufficiently small steps and using the approximation (\ref{pav2}) only for $E_k \geq E_{\hbox{\scriptsize av}}$, while for $E_k < E_{\hbox{\scriptsize av}}$ using the full formula (\ref{Vkpk2-App}), where at each step the integral would require only the knowledge of $\lambda[E]$ for $E > E_{\hbox{\scriptsize av}}$. In principle, that integral may extend to $E >0 $, but this corresponds to sufficiently large values of $p_k$, for which $V_k(p_k)$ is guaranteed to exhibit fast exponential decay. Therefore, a cuttoff
$p_{kC}$ can be imposed that does not allow the upper integration limit to extend above $0$. This cutoff is further discussed in Appendix~\ref{condensation}.

\section{System of many spins 1/2}
\label{spins}

In order to appreciate certain general aspects of the macroscopic case, it is sufficient to consider an otherwise very artificial  example of $N_s \gg 1$ non-interacting spins 1/2 in magnetic field. 
The number of levels in this system is
\begin{equation}
 N = 2^{N_s} \gg N_s.
\label{2Ns}
\end{equation}

In the basis of spins quantized along the direction of the magnetic field ($z$-direction), each eigenstate is determined by a set of spin projections $S_{nz} = \pm 1/2 $, and the corresponding energy is 
\begin{equation}
 E = \sum_{n=1}^{N_s} \Omega S_{nz}
\label{espin}
\end{equation}
where $\Omega$ is the ``Larmor energy'' associated with the splitting of spin states. The minimum and the maximum energies of this spectrum are respectively:
\begin{equation}
 E_{\hbox{\scriptsize min}} = - {1 \over 2} N_s  \Omega,
\label{esmin}
\end{equation}
and
\begin{equation}
 E_{\hbox{\scriptsize max}} = {1 \over 2} N_s  \Omega.
\label{esmax}
\end{equation}
The average of all energies in the spectrum is zero, and the mean-squared deviation from this average is
\begin{equation}
 \sigma_s = {1 \over 4} N_s  \Omega^2.
\label{sigma-s-App}
\end{equation}
All energies have form 
\begin{equation}
 E = E_{\hbox{\scriptsize min}} + \Omega m,
\label{eslevels}
\end{equation}
where $m$ is an integer number between $0$ and $N_s$.
The degeneracy $D_m$ of the energy levels corresponding to a given value of $m$ is
\begin{equation}
 D_m = {N_s! \over m! (N_s - m)!}.
\label{Dm}
\end{equation}
For $m, (N_s - m) \gg 1$, one can use the Stirling approximation 
$m! \approx \sqrt{2 \pi m} e^{m (\hbox{log} m - 1)}$ and likewise for 
$(N_s - m)!$ and $N_s!$, and then divide the degeneracy $D_m$ by distance between levels $\Omega$ to obtain the density of states:
\begin{equation}
 \nu(E) = {1 \over \Omega} \sqrt{{N_s \over 2 \pi m (N_s - m)} }
\ \ e^{N_s \hbox{log}{N_s \over N_s - m} + m \hbox{log}{N_s - m \over m}},
\label{nuStir}
\end{equation}
where $m = (E -E_{\hbox{\scriptsize min}})/\Omega $ --- as follows from (\ref{eslevels}). 
The density of states can be further approximated by Gaussian:
\begin{equation}
 \nu(E) = {N \over \sqrt{2 \pi \sigma_S} }
e^{-{E^2 \over 2 \sigma_s}},
\label{nuGauss}
\end{equation}
which is  the consequence of the Central Limit Theorem. It becomes increasingly inaccurate for $|E| \gg \sqrt{\sigma_s}$.

As the most representative physical situation, let us consider the case $T = \Omega$, where $T$ is the usual temperature defined by Eq.(\ref{T}). In this case, the Gaussian approximation (\ref{nuGauss}) for would imply $E_{\hbox{\scriptsize av}} = - {1 \over 4} \Omega N_s = E_{\hbox{\scriptsize min}}/2$.  This approximation is not very accurate for $E_{\hbox{\scriptsize av}} \sim E_{\hbox{\scriptsize min}}$. For example, the accurate approximation (\ref{nuStir}) gives $T = \Omega/\hbox{log}3$ for $E_{\hbox{\scriptsize av}} = E_{\hbox{\scriptsize min}}/2$. Yet this discrepancy does not invalidate the general estimate that
in the physical range of interest
\begin{equation}
 |E_{\hbox{\scriptsize av}}| \gg \sqrt{\sigma_s},
\label{eavss}
\end{equation}
with most routine being the case when $E_{\hbox{\scriptsize av}}$ is negative and a finite fraction of $E_{\hbox{\scriptsize min}}$.

Now, I would like to show that condition 
(\ref{em-el}), which guarantees the smallness of all $p_k$, is violated for the present spin system almost immediately after $E_{\hbox{\scriptsize av}}$ starts decreasing below zero.  Let us consider what it takes to satisfy Eq.(\ref{nueav}), when condition (\ref{em-el}) is fulfilled. The integral in this equation has a character of the average value of $(E - E_{\hbox{\scriptsize av}})$ under the effective distribution described by a product of slow varying function $1/[1+ \lambda (E - E_{\hbox{\scriptsize av}})]$ and sharply peaked symmetric function $\nu(E)$, which for the present purpose is well approximated by the Gaussian (\ref{nuGauss}).  In the leading order, this effective distribution will remain symmetric but with respect to  a maximum, which is slightly shifted relatively to that of $\nu(E)$. Equation(\ref{nueav}) can only be satisfied, when that maximum  coincides with $E_{\hbox{\scriptsize av}}$.  Such a condition gives $\lambda = E_{\hbox{\scriptsize av}}/\sigma_s$. Then requiring that $E_{\lambda} = E_{\hbox{\scriptsize min}}$, I obtain the critical value of $E_{\hbox{\scriptsize av}}$  denoted earlier [Eq.(\ref{ElEC})] as $E_C$:
\begin{equation}
 E_C = {\sigma_s \over E_{\hbox{\scriptsize min}}} = -{1 \over 2} \Omega \ll \sigma_s,
\label{EC}
\end{equation}
which implies that the small-$p_k$ condition (\ref{em-el}) is violated at least for some $k$, when the values of $E_{\hbox{\scriptsize av}}$ fall in the range of primary physical interest (\ref{eavss}).

The analysis in Section~\ref{macro} relies on the assumption that, for $N_s \sim 10^{23}$,
\begin{equation}
 E_2 - E_{\hbox{\scriptsize min}} \gg {|E_{\hbox{\scriptsize min}}| \over N} ,
\label{eps2}
\end{equation}
where $E_2$ is the energy of the second lowest level, 
$E_{\hbox{\scriptsize av}}$ satisfies (\ref{eavss}).
The above property is obviously true for the present spin system, where
$E_2 - E_{\hbox{\scriptsize min}} = \Omega$, but it should also remain valid for any
realistic macroscopic system with non-degenerate ground state ---consequence of the fact (\ref{2Ns}) that $N$ is exponentially larger than $N_s$.

Indeed, let us consider as a more realistic example, the system of $N_s$ spins 1/2 on a cubic lattice with ferromagnetic nearest-neighbor exchange interaction and periodic boundary conditions. The Hamiltonian is ${\cal H} =
- J \sum_{\hbox{\tiny NN}} \mathbf S_m \cdot \mathbf S_n$, where $\mathbf S_m$ are vector spin operators, $J$ is the (positive)  exchange  constant, and NN refers to the summation over the nearest neighbor $\{m,n\}$-pairs. In this case, the ground state has energy $E_{\hbox{\scriptsize min}} = - {3 \over 4} J N_s$. It is $N_s$-fold degenerate, because all states with total spin ${1\over 2} N_s$ have this energy. The excited states have a character of spin waves.  Yet these spin waves are gapped by the finite linear size of the system ($^3\sqrt{N_s}$), i.e. $E_2 - E_{\hbox{\scriptsize min}} \sim {J \over ^3\sqrt{N_s}} \gg {|E_{\hbox{\scriptsize min}}| \over N}$. As far as the ground state degeneracy is concerned, one can imagine, that any, even unrealistically small stray magnetic field would lift this degeneracy with the energy splitting that would still be exponentially larger than ${|E_{\hbox{\scriptsize min}}| \over N}$. (It is also not difficult to extend the result of this work to the theoretical case of degenerate ground state.)

\section{Condensation into the lowest level of macroscopic system}
\label{condensation}

When condition (\ref{emin-cond}) is not fulfilled, volume $V_k(p_k)$ does not decay exponentially fast. The possible alternatives are (i) that it changes slowly ---decreases or increases, or (ii) it  has a sharp Gaussian-like maximum
at $p_k = p_{k0}$ somewhere between 0 and 1,  in which case  
$ \langle p_k \rangle \approx p_{k0}$.
In the both cases, $\langle p_k \rangle$ would be a significant fraction of 1.

The first of the above cases might be realized in numerical studies, when $N \gg 1$, but log$N$ is not too large. In this case, several low-lying levels may exhibit large values of $\langle p_k \rangle$.

I now show that in the macroscopic system, of the type described in Section\ref{macro}, it is the second of the above cases that is realized, and that the significant average occupation builds only for the lowest-lying level, while for all other levels, continue exhibiting occupations $\langle p_k \rangle \ll 1$ describable by formula (\ref{pav2}). 

The subtlety of the present part is that it requires one to find a correction to a very accurate approximation (\ref{em-el}). In doing so, one needs to go back and forth between the scale of $|E_{\hbox{\scriptsize min}}|$ and the scale of the level spacing between the lowest two levels.

I begin with several general observations, which are applicable to both cases (i) and (ii).

1) Let us denote as $I_1$, the sum over the populations of those exceptional low-lying levels not describable by formulas (\ref{Vkpk1},\ref{pav2}) and exhibiting large average occupations $\langle p_k \rangle$.
As mentioned in Section~\ref{beyond}, there can exist only a relatively small number of these levels. Therefore, their energies should all be close to $E_{\hbox{\scriptsize min}}$.

All other levels --- in particular those surrounding the dominant peak in the density of states $\nu(E)$ --- would have ``regular'' values of $\langle p_k \rangle$ given by  formula (\ref{pav2}). Given the overall normalization (\ref{normav}), the total occupation of regular levels, to be denoted as $I_2$, is 
\begin{equation}
 I_2 = 1 - I_1.
\label{I2-I1}
\end{equation}

I can now use the energy condition (\ref{epsavav}) and the fact that the occupations of the exceptional levels are peaked around $E_{\hbox{\scriptsize min}}$, while the integrand ${\nu(E) \over 1 - \lambda (E - E_{\hbox{\scriptsize av}})}$ is also peaked around $E_C \approx 0$ [see Eq.(\ref{EC}) and its derivation], to obtain 
\begin{equation}
 (E_{\hbox{\scriptsize min}} - E_{\hbox{\scriptsize av}}) I_1 - {E_{\hbox{\scriptsize av}} \over 1 - \lambda E_{\hbox{\scriptsize av}}} I_2 = 0,
\label{I1I2}
\end{equation}
which, in combination with (\ref{I2-I1}) and the approximation (\ref{lambdaA}) for $\lambda$ gives
\begin{equation}
 I_1 \approx {E_{\hbox{\scriptsize av}} \over E_{\hbox{\scriptsize min}}},
\label{I1}
\end{equation}
\begin{equation}
 I_2 \approx {E_{\hbox{\scriptsize min}} - E_{\hbox{\scriptsize av}} \over E_{\hbox{\scriptsize min}}}.
\label{I2}
\end{equation}

\

2) When the general formula (\ref{Vkpk2}) is used, one has to remember that manifold $M_k$ exists and its volume $V_k$ has non-zero value, only when 
$E_{\hbox{\scriptsize min}}^{\prime} \leq E_{\hbox{\scriptsize av}} \leq E_{\hbox{\scriptsize max}}^{\prime}$, where $E_{\hbox{\scriptsize min}}^{\prime}$ and $E_{\hbox{\scriptsize max}}^{\prime}$ are the minimum and the maximum value among all energies excluding $E_k$.
This means that the argument of $\lambda[E]$ should also stay within the same limits, which, in turn imposes cutoff on $V_k(p_k)$ as a function of $p_k$, when the upper integration limit in Eq.(\ref{Vkpk2}) reaches $E_{\hbox{\scriptsize max}}^{\prime}$ or $E_{\hbox{\scriptsize min}}^{\prime}$. Beyond this cuttoff, $V_k(p_k) = 0$.

Here I am primarily concerned with the exceptional low-lying levels (and also exclude the case $E_{\hbox{\scriptsize av}} < E_2$), which translates into constraint
\begin{equation}
 E_{\hbox{\scriptsize av}} - { (E_k - E_{\hbox{\scriptsize av}}) p_k \over 1 - p_k}  \leq E_{\hbox{\scriptsize max}}.
\label{Emax-limit}
\end{equation}
This results in the upper maximum value for $p_k$:
\begin{equation}
 p_{k[\hbox{\scriptsize max}]} = {E_{\hbox{\scriptsize max}} - E_{\hbox{\scriptsize av}} \over E_{\hbox{\scriptsize max}} - E_k}.
\label{pkmax}
\end{equation}

One can establish a stronger effective cutoff for $p_k$, beyond which $V_k(p_k)$ is guaranteed to exhibit sharp exponential decay (for $E_k < E_{\hbox{\scriptsize av}}$). This cutoff corresponds to 
\begin{equation}
 E_{\hbox{\scriptsize av}} - { (E_k - E_{\hbox{\scriptsize av}}) p_k \over 1 - p_k}  =0,
\label{0-limit}
\end{equation}
which translates into the cuttoff value
\begin{equation}
 p_{kC} = {E_{\hbox{\scriptsize av}} \over E_k}.
\label{pkC}
\end{equation}

The above cuttoff originates from the following argument. When $V_k(p_k)$ given by (\ref{Vkpk2})  decays slowly or increases at small $p_k$, it happens, because the increasing value of integral in (\ref{Vkpk2}) nearly compensates or outweighs the decreasing value of the preceeding logarithmic term.  I note now, that $\lambda[E]$ becomes negative above $E=0$. Therefore, once $p_k$ reaches the value given by Eq.(\ref{0-limit}), the integration extends into positive $E$ and the integral starts decreasing. After that, it is certain that nothing any longer can slow the fast exponential decay --- hence cutoff (\ref{pkC}).

\

3) Cutoff (\ref{pkC}) can now be used to show that as $E_{\hbox{\scriptsize av}}$ decreases, the value  of $E_{\lambda}[E_{\hbox{\scriptsize av}}]$ cannot always stay below $E_{\hbox{\scriptsize min}}$. In principle, the estimate (\ref{EC}) already indicates this, but it could have happened that it signified only $E_{\lambda}$  approaching $E_{\hbox{\scriptsize min}}$ within $O(1/N)$ from below and never crossing it.
 
If the latter possibility were to be realized, the derivative of $V_1(p_1)$ would remain negative or zero for $0 \leq p_1 \leq 1$ with rapid exponential drop above $p_{1C}$ given by Eq.(\ref{pkC}). This would, in turn, imply that 
\begin{equation}
 \langle p_1 \rangle < {1 \over 2} p_{1C} + O(1/N) 
= {1 \over 2} {E_{\hbox{\scriptsize av}} \over E_{\hbox{\scriptsize min}}} + O(1/N).
\label{p1av-p1C}
\end{equation}
Given inequality (\ref{eps2-m}), $\langle p_2 \rangle \ll 1$ for $E_{\lambda} < E_{\hbox{\scriptsize min}}$. Therefore, the only ``exceptional'' level in this case would be the lowest one. That is, $I_1 = \langle p_1 \rangle$, where $\langle p_1 \rangle$ is given by Eq.(\ref{p1av-p1C}), which contradicts to Eq.(\ref{I1}). Thus the estimate (\ref{EC}) indicated correctly that $E_{\lambda}$ crossing $E_{\hbox{\scriptsize min}}$ corresponds to $E_{\hbox{\scriptsize av}} = E_C \ll E_{\hbox{\scriptsize min}}$. 

\

4) When $E_{\lambda} > E_{\hbox{\scriptsize min}}$, the volume $V_1(p_1)$ increases at small $p_1$; then it reaches a maximum at the value of $p_1$ to be denoted at $p_{10}$, and, finally decays to zero at $p_1 = 1$.

Below I locate $p_{10}$ and show that it corresponds to a $\delta$-function-like maximum:
\begin{equation}
 V_1(p_1) \sim \hbox{exp} \left[ - {(p_1 - p_{10})^2 \over 2 \sigma_1} \right],
\label{V1max}
\end{equation}
where $\sqrt{\sigma_1} \ll 1$, which implies 
\begin{equation}
 \langle p_1 \rangle \approx p_{10}.
\label{p10}
\end{equation}

Requiring the derivative of the exponential power in Eq.(\ref{Vkpk2}) with respect to $p_k$ to be equal to zero, I obtain for $k=1$:
\begin{equation}
 1 + \lambda \left[ E_{\hbox{\scriptsize av}} - {(E_{\hbox{\scriptsize min}} - E_{\hbox{\scriptsize av}}) p_{10} \over 1 - p_{10}} \right] 
{E_{\hbox{\scriptsize min}} - E_{\hbox{\scriptsize av}} \over 1 - p_{10}} = 0.
\label{Vk-deriv}
\end{equation}
Making in (\ref{Vk-deriv}) substitution
\begin{equation}
 \lambda[E] = {1 \over E - E_{\lambda}[E]},
\label{l-el}
\end{equation}
I obtain after some manipulation:
\begin{equation}
 E_{\lambda}\left[ E_{\hbox{\scriptsize av}} - {(E_{\hbox{\scriptsize min}} - E_{\hbox{\scriptsize av}}) p_{10} \over 1 - p_{10}} \right] = E_{\hbox{\scriptsize min}}
\label{el-ec-cond}
\end{equation}
Therefore, according to the definition of $E_C$ given by Eq.(\ref{ElEC}),
\begin{equation}
 E_{\hbox{\scriptsize av}} - {(E_{\hbox{\scriptsize min}} - E_{\hbox{\scriptsize av}}) p_{10} \over 1 - p_{10}} = E_C.
\label{p10-ec -cond}
\end{equation}
As a result,
\begin{equation}
 p_{10} = {E_C - E_{\hbox{\scriptsize av}} \over E_C - E_{\hbox{\scriptsize min}}} \approx {E_{\hbox{\scriptsize av}} \over E_{\hbox{\scriptsize min}}}.
\label{p10-result}
\end{equation}
The transition to the approximate value above follows from inequality (\ref{EC-m}). 

The approximate value in (\ref{p10-result}) is equal to the effective cutoff value $p_{1C}$ given by (\ref{pkC}), where the derivative of $V_k(p_k)$ is supposed to be strongly negative. There is, however, no contradiction here.
The accurate value of $p_{10}$ given by the middle expression in Eq.(\ref{p10-result}) is smaller than $p_{1C}$, and, as shown below, the width of the maximum $\sqrt{\sigma_1}$ is much smaller than the difference between $p_{1C}$ and $p_{10}$.

The value of $\sigma_1$ can now be obtained from
\begin{equation}
 {1 \over \sigma_1} = 
- \left. { d^2 \hbox{log} V_1(p_1) \over d p_1^2} \right\vert_{p_1 = p_{10}}
= - {(N-3) \over (1-p_{10})^2}
 \left. {d E_{\lambda}[E] \over dE} \right\vert_{E = E_C} .
\label{sigma1}
\end{equation}
Condition $-\left. {d E_{\lambda}[E] \over dE} \right\vert_{E = E_C} \gg 1/N$
would then be sufficient to justify $\sigma_1 \ll 1$. 

By differentiating the accurate version of conditions (\ref{I2-I1},\ref{I1I2}) and using formula (\ref{Vkpk2}) for $V_1(p_1)$ to calculate $I_1 = \langle p_1 \rangle$, I was able to obtain
\begin{equation}
 -\left. {d E_{\lambda}[E] \over dE} \right\vert_{E = E_C} \approx 
{2 \over \pi} {E_{\hbox{\scriptsize min}}^2 \over \sigma_s} \sim N_s \gg 1,
\label{el-deriv}
\end{equation}
which guarantees the validity of Eq.(\ref{p1av-p1C}) with $p_{10}$ approximated by Eq.(\ref{p10-result}) ---hence
\begin{equation}
 \langle p_1 \rangle \approx {E_{\hbox{\scriptsize av}} \over E_{\hbox{\scriptsize min}}}.
\label{p1av-App}
\end{equation}

Finally, I observe, that  Eqs.(\ref{I1}) and (\ref{p1av-App}) imply that 
\begin{equation}
 \langle p_1 \rangle \approx I_1.
\label{p1I1}
\end{equation}
The approximate values (\ref{I1}) and (\ref{p1av-App}) for $I_1$ and $\langle p_1 \rangle$ were obtained under the same assumption $E_C \approx 0$. In fact, Eq.(\ref{p1I1}) also holds, when $E_C$ is not neglected.

Equation (\ref{p1I1}) indicates that $\langle p_1 \rangle$ exhausts or almost exhausts the total occupation $I_1$ of all exceptional levels, whose respective Hilbert space volumes depart significantly from formula (\ref{Vkpk1}). 

One can further rule out the possibility ``almost exhausts''. It would require that $E_{\lambda}[E_{\hbox{\scriptsize av}}]$ reaches at least the second lowest energy level $E_2$. If this
were to happen at $E_{\hbox{\scriptsize av}} = E_{C2}$, then $V_2(p_2)$ would aquire maximum at 
\begin{equation}
 p_{20} = {E_{C2} - E_{\hbox{\scriptsize av}} \over E_{C2} - E_{\hbox{\scriptsize min}}}
\label{p20}
\end{equation}
for all $E_{\hbox{\scriptsize av}} < E_{C2}$. This maximum, sharp or not, would then lead to $\langle p_1 \rangle + \langle p_2 \rangle$ becoming significantly greater than $E_{\hbox{\scriptsize av}} / E_{\hbox{\scriptsize min}}$ for most of the range $E_{\hbox{\scriptsize av}} < E_{C2}$ in contradiction with Eq.(\ref{I1}).

Therefore, $E_{\lambda}$ is always sufficently smaller than $E_2$ to justify the use of formula (\ref{pav2}) for all $k \geq 2$. 

\section{Derivation of Eqs. (\ref{rho1-av}-\ref{rhoa-avA}) for the density matrix of a subsystem in a macroscopic environment}
\label{subsystem-App}

It follows from inequality (\ref{EC-m}) that $E_{\lambda}[E]$ has the following approximate property:
\begin{equation}
 \hbox{if} \ \ E_{\lambda}[E] < E_{\hbox{\scriptsize min}}, \ \hbox{then} \ E \approx E_P,
\label{Ec-cond1}
\end{equation}
and
\begin{equation}
 \hbox{if} \ \ E_{\hbox{\scriptsize min}} < E < E_P, \ \hbox{then} \ E_{\lambda}[E] \approx E_{\hbox{\scriptsize min}},
\label{Ec-cond2}
\end{equation}
where $E_P$ is the position of the sharp maximum of $\nu(E)$.
It was also obtained in Appendix~\ref{condensation} that
\begin{equation}
 E_{\lambda}[E] < E_2, \ \ \ \ \forall E < E_P.
\label{El-E2}
\end{equation}
where $E_2$ is the energy of the second lowest level.

The key assumption of the present calculation is that the above conditions apply  to both sets of energies $\{ E_a \}$ and $\{ E_b \}$ with the appropriate insertion of subscripts $a$ and $b$. The densities of states $\nu_a(E)$ and $\nu_b(E)$ corresponding to these two sets are characterized by the respective peak positions:
\begin{equation}
 E_{Pa} = E_{S \alpha},
\label{EPa}
\end{equation}
\begin{equation}
 E_{Pb} = -{ E_{S \alpha} \over N_1 -1}
\label{EPb}
\end{equation}

Each energy set also has a respective minimum value $E_{a \hbox{\scriptsize min}}$ and $E_{b \hbox{\scriptsize min}}$
The outcome of the calculation now simply depends on whether $E_{a \hbox{\scriptsize min}}$ is smaller or larger than $E_{b \hbox{\scriptsize min}}$.

Case I: $\alpha \geq 2$

In this case:
\begin{equation}
 E_{a \hbox{\scriptsize min}} = E_{S \alpha} + E_{E \hbox{\scriptsize min}},
\label{EaminI}
\end{equation}
and 
\begin{equation}
 E_{b \hbox{\scriptsize min}} = E_{S \hbox{\scriptsize min}} + E_{E \hbox{\scriptsize min}} \equiv E_{\hbox{\scriptsize min}},
\label{EbminI}
\end{equation}
which implies that $E_{a \hbox{\scriptsize min}} > E_{b \hbox{\scriptsize min}}$. 

Given conditions (\ref{El-E2}) and (\ref{l-space}), I can further constrain
\begin{equation}
 E_{\lambda_b} < E_{b2} \lesssim E_{a \hbox{\scriptsize min}}
\label{Elb-Eb2-Eamin}
\end{equation}
This condition applies to $E_{\lambda_b}[E]$ independently of its argument. Therefore,
\begin{equation}
 E_{\lambda_a} \left[ E_{\hbox{\scriptsize av}} + {E_A \over \rho_{\alpha}} \right] < E_{a \hbox{\scriptsize min}}.
\label{Elb-Eamin}
\end{equation}
Thus, according to (\ref{Ec-cond1}),
\begin{equation}
 E_{\hbox{\scriptsize av}} + {E_A \over \rho_{\alpha}} \approx E_{Pa} = E_{S \alpha}.
\label{eq1-I}
\end{equation}
One can now solve the pair of equations (\ref{EA},\ref{eq1-I}) with respect to $E_A$ and $\rho_{\alpha}$ to obtain
\begin{equation}
 \rho_{\alpha} = {1 \over N_1} 
{
E_{\hbox{\scriptsize av}} - E_{\lambda_a} \left[ E_{\hbox{\scriptsize av}} + {E_A \over \rho_{\alpha}} \right]
\over
E_{S \alpha} - E_{\lambda_a} \left[ E_{\hbox{\scriptsize av}} + {E_A \over \rho_{\alpha}} \right]
}.
\label{rhoalphaI-tmp}
\end{equation}
Now, in order to find 
$E_{\lambda_a} \left[ E_{\hbox{\scriptsize av}} + {E_A \over \rho_{\alpha}} \right]$,
one is helped by the inequality
\begin{equation}
 E_{\hbox{\scriptsize av}} < E_{Pa}, E_{Pb},
\label{Eav-Epa}
\end{equation}
which is the consequence of the earlier assumption (\ref{s2A}) that $|E_{\hbox{\scriptsize av}}| \gg E_{S \alpha}, \ \forall \alpha$. Equations (\ref{eq1-I},\ref{Eav-Epa}) now require that $E_A > 0$. As a result,
\begin{equation}
 E_{\hbox{\scriptsize av}} - {E_A \over 1- \rho_{\alpha}} < E_{\hbox{\scriptsize av}} < E_{Pb},
\label{ineq1-I}
\end{equation}
which, according to (\ref{Ec-cond2}), implies that
\begin{equation}
 E_{\lambda_b} \left[ 
E_{\hbox{\scriptsize av}} - {E_A \over 1- \rho_{\alpha}}
\right] = E_{b \hbox{\scriptsize min}}.
\label{Eav-Ebmin}
\end{equation}
Therefore, according (\ref{Ela-Elb}) and (\ref{EbminI}). 
\begin{equation}
 E_{\lambda_a} \left[ 
E_{\hbox{\scriptsize av}} + {E_A \over \rho_{\alpha}}
\right] = E_{\hbox{\scriptsize min}}.
\label{EaminI-A}
\end{equation}
The substitution of Eq.(\ref{EaminI-A}) into Eq.(\ref{rhoalphaI-tmp}) yields
\begin{equation}
 \rho_{\alpha} = {1 \over N_1} 
{
E_{\hbox{\scriptsize av}} - E_{\hbox{\scriptsize min}} 
\over
E_{S \alpha} - E_{\hbox{\scriptsize min}} }.
\label{rhoalphaI}
\end{equation}

Given inequality (\ref{s2A}), the leading order approximation of (\ref{rhoalphaI}) is 
\begin{equation}
 \rho_{\alpha} = 
{1 \over N_1} \
\left( 1 - 
{ E_{\hbox{\scriptsize av}}  \over  E_{E \hbox{\scriptsize min}} } 
\right).
\label{rhoalphaI-A}
\end{equation}

\


Case II: $\alpha =1$.

Now 
\begin{equation}
 E_{a \hbox{\scriptsize min}} = E_{S \hbox{\scriptsize min}} + E_{E \hbox{\scriptsize min}} \equiv E_{\hbox{\scriptsize min}},
\label{EaminII}
\end{equation}
and 
\begin{equation}
 E_{b \hbox{\scriptsize min}} = E_{S2} + E_{E \hbox{\scriptsize min}} ,
\label{EbminII}
\end{equation}
implying $E_{a \hbox{\scriptsize min}} < E_{b \hbox{\scriptsize min}}$. 

One can obtain using condition (\ref{l-space}), that $E_{b \hbox{\scriptsize min}} - E_{a \hbox{\scriptsize min}} = E_{S2} - E_{S \hbox{\scriptsize min}} \geq E_{E2} - E_{E \hbox{\scriptsize min}} = E_{a2} - E_{a \hbox{\scriptsize min}}$. The resulting inequality together with condition (\ref{El-E2}) implies that 
\begin{equation}
 E_{\lambda_a} < E_{a2} \leq E_{b \hbox{\scriptsize min}}.
\label{Ela-Ebmin}
\end{equation}
Inequality (\ref{Ela-Ebmin}) applies to any argument of $E_{\lambda_a}[E]$ including
$E_{\hbox{\scriptsize av}} + {E_A \over \rho_1}$.  Using this fact together with 
Eq.(\ref{Ela-Elb}), I obtain
\begin{equation}
 E_{\lambda_b} \left[ E_{\hbox{\scriptsize av}} - {E_A \over 1 - \rho_1} \right] < E_{b \hbox{\scriptsize min}}.
\label{Elb-Ebmin}
\end{equation}
Thus, according to (\ref{Ec-cond1}),
\begin{equation}
 E_{\hbox{\scriptsize av}} - {E_A \over 1 - \rho_1} \approx E_{Pb} 
= -{ E_{S \hbox{\scriptsize min}} \over N_1 - 1}.
\label{eq1-II}
\end{equation}

Given inequality (\ref{Eav-Epa}), Equation (\ref{eq1-II}) implies that 
$E_A < 0$. As a result,
\begin{equation}
 E_{\hbox{\scriptsize av}} + {E_A \over \rho_1} < E_{\hbox{\scriptsize av}} < E_{Pa},
\label{ineq1-II}
\end{equation}
and thus, according to (\ref{Ec-cond2}), 
\begin{equation}
 E_{\lambda_a} \left[ 
E_{\hbox{\scriptsize av}} + {E_A \over  \rho_1 }
\right] = E_{\hbox{\scriptsize min}}.
\label{Eav-Emin}
\end{equation}

After Eq.(\ref{Eav-Emin}) is substituted into (\ref{EA}) and the resulting equation is solved jointly with (\ref{eq1-II}) with respect to $\rho_1$ and $E_A$ , I obtain
\begin{equation}
 \rho_1 =  
{
E_{\hbox{\scriptsize av}} \left( 1- {1 \over N_1} \right) +  
{E_{S \hbox{\scriptsize min}} \over N_1 -1} +
{E_{\hbox{\scriptsize min}} \over N_1}
\over
{E_{S \hbox{\scriptsize min}} \over N_1 -1} +  E_{\hbox{\scriptsize min}} }.
\label{rho1II}
\end{equation}

Given the condition $|E_{S \hbox{\scriptsize min}}| \ll |E_{\hbox{\scriptsize av}}|, |E_{E \hbox{\scriptsize min}}|$, the leading order approximation of (\ref{rho1II}) is 
\begin{equation}
 \rho_1 = { E_{\hbox{\scriptsize av}}  \over  E_{E \hbox{\scriptsize min}} } 
+
{1 \over N_1} \
\left( 1 - 
{ E_{\hbox{\scriptsize av}}  \over  E_{E \hbox{\scriptsize min}} } 
\right).
\label{rho1II-A}
\end{equation}

\

It is clear from the derivation that formulas (\ref{rhoalphaI},\ref{rho1II}) describe a broader range of mostly abstract possibilities beyond the physical assumptions used so far. 

One can, in particular, consider relaxing the condition $|E_{S \alpha}| \ll |E_{\hbox{\scriptsize av}}|, |E_{E \hbox{\scriptsize min}}|$. In this case, however, on needs to be concerned with the validity of conditions (\ref{Ec-cond1}, \ref{Ec-cond2}) for the energy set $\{E_b\}$. For example, if the spectrum of $\{ E_{S \alpha} \}$ consists of only a few far separated levels, then $\nu_b(E)$ loses the single peak structure and hence conditions  (\ref{Ec-cond1}, \ref{Ec-cond2}). However, when $N_1 \gg 1$, and the spectrum of energies $\{ E_{S \alpha} \}$ has a sharply-peaked Gaussian-like density of states, then conditions (\ref{Ec-cond1}, \ref{Ec-cond2}) are recovered.

In the latter case, one also needs to examine what happens, when $E_{\hbox{\scriptsize av}}$ falls between $E_{Pa}$ and $E_{Pb}$.  In this case, the system of equations (\ref{EA}, \ref{Ela-Elb}) with approximations (\ref{Ec-cond1}, \ref{Ec-cond2})
generates two more solutions, in addition to the one already found. These two solutions correspond to the possibility that both 
$E_{\hbox{\scriptsize av}} + {E_A \over \rho_{\alpha}}$ and $E_{\hbox{\scriptsize av}} - {E_A \over 1 - \rho_{\alpha}}$ become approximately equal to $E_{Pa}$ and $E_{Pb}$ respectively. One solution would then correspond to positive values and the other one to the negative values of $\lambda_a$, $\lambda_b$. They are presumably a maximum and a minimum located close to each other in the plane of variables $(E_A, \rho_{\alpha})$. The additional maximum should, presumably give a smaller Hilbert space maximum, in comparison to the one found earlier, because it has a larger value of $\rho_{\alpha}$, and therefore stronger suppressed by term containing log$(1- \rho_{\alpha})$ in Eq.(\ref{V-rho}).

\end{document}